\documentclass[aps,prd,10pt,twocolumn, superscriptaddress, nofootinbib]{revtex4}
\usepackage{mathrsfs}  
\usepackage{epsfig, cancel}
\usepackage{latexsym}
\usepackage{natbib, comment}
\usepackage{url}
\usepackage{dcolumn}
\usepackage{multirow}
\usepackage{color}
\usepackage{cancel}
\usepackage{soul}
\usepackage[normalem]{ulem}
\usepackage{amsfonts,amssymb,amsmath, txfonts}
\usepackage{graphicx,epsfig}
\usepackage{psfrag}
\usepackage{subfigure}
\usepackage{hyperref}
\hypersetup{colorlinks=true}
\usepackage{mathtools}
\usepackage{enumitem}
\usepackage{float}
\usepackage[dvipsnames]{xcolor}
\usepackage{xcolor}
\usepackage{longtable}
\hypersetup{ linktoc=all, 
    colorlinks, linkcolor={red}, 
    citecolor={blue}, urlcolor={blue}
}
\definecolor{rosy}{RGB}{230,235,252}
\definecolor{myframetitle}{RGB}{90,89,170}
\definecolor{myblocktitle}{RGB}{140,185,249}
\definecolor{mytitle}{RGB}{10,80,26}

\definecolor{darkgreen}{RGB}{27,130,45}
\definecolor{darkblue}{rgb}{0,0,0.3}
\definecolor{darkred}{rgb}{0.7,0,0}

\definecolor{light gray}{RGB}{220,220,220}
\definecolor{dark purple}{RGB}{108,0,217}
\definecolor{pink}{RGB}{190,20,100}
\definecolor{orang}{RGB}{193,63,0}
\definecolor{green}{RGB}{11,98,17}
\definecolor{darkpink}{RGB}{153,0,76}
\definecolor{bluegreen}{RGB}{0,102,102}
\definecolor{greenlagan}{RGB}{0,102,0}
\definecolor{redgreen}{RGB}{102,102,0}
\definecolor{Redgreen}{RGB}{153,76,0}
\definecolor{vividviolet}{rgb}{0.62, 0.0, 1.0}
\definecolor{amaranth}{rgb}{0.9, 0.17, 0.31}
\definecolor{palatinateblue}{rgb}{0.15, 0.23, 0.89}
\definecolor{brightpink}{rgb}{1.0, 0.0, 0.5}
\definecolor{cornflowerblue}{rgb}{0.39, 0.58, 0.93}
\definecolor{deepcarminepink}{rgb}{0.94, 0.19, 0.22}
\definecolor{radicalred}{rgb}{1.0, 0.21, 0.37}
\def\green{\textcolor{green}}
\def\H0{{\text{H}\hspace*{-2.05mm}\text{H} 0\hspace*{-1.35mm}0\ }}

\def\be{\begin{equation}}
\def\ee{\end{equation}}
\def\beq{\begin{equation}}
\def\eeq{\end{equation}}
\def\bea{\begin{eqnarray}}
\def\eea{\end{eqnarray}}

\begin{document}

\title{On Larger $H_0$ Values in the CMB Dipole Direction}

\author{Orlando Luongo}\email{orlando.luongo@unicam.it}
\affiliation{%
Dipartimento di Matematica, Universit\`a di Pisa, Largo B. Pontecorvo 5, Pisa, 56127, Italy.}
\affiliation{%
Divisione di Fisica, Universit\`a di Camerino, Via Madonna delle Carceri, 9, 62032, Italy.}
\affiliation{%
Al-Farabi Kazakh National University, Al-Farabi av. 71, 050040 Almaty, Kazakhstan.}
\affiliation{National Nanotechnology Laboratory of Open Type,  Almaty 050040, Kazakhstan.}

\author{Marco Muccino} \email{muccino@lnf.infn.it}
\affiliation{%
Al-Farabi Kazakh National University, Al-Farabi av. 71, 050040 Almaty, Kazakhstan.}
\affiliation{National Nanotechnology Laboratory of Open Type,  Almaty 050040, Kazakhstan.}
\affiliation{Istituto Nazionale di Fisica Nucleare (INFN), Laboratori Nazionali di Frascati, 00044 Frascati, Italy.}

\author{Eoin \'O Colg\'ain}\email{eoin@sogang.ac.kr}
\affiliation{Center for Quantum Spacetime, Sogang University, Seoul 121-742, Korea.}
\affiliation{Department of Physics, Sogang University, Seoul 121-742, Korea.} 
\author{M. M. Sheikh-Jabbari}\email{shahin.s.jabbari@gmail.com}
\affiliation{School of Physics, Institute for Research in Fundamental Sciences (IPM), P.O.Box 19395-5531, Tehran, Iran.}
\author{Lu Yin}\email{yinlu@sogang.ac.kr}
\affiliation{Center for Quantum Spacetime, Sogang University, Seoul 121-742, Korea.}
 \affiliation{Department of Physics, Sogang University, Seoul 121-742, Korea.}

\begin{abstract}
On the assumption that quasars (QSO) and gamma-ray bursts (GRB) represent \textit{standardisable candles}, we provide evidence that the Hubble constant $H_0$ adopts larger values in hemispheres aligned with the CMB dipole direction. If substantiated, this trend signals a departure from FLRW cosmology. In particular, QSOs show a definite trend, whereas our findings in GRBs are consistent with an isotropic Universe, but we show in a sample of GRBs calibrated with Type Ia supernovae (SN) that this conclusion may change as one focuses on GRBs more closely (mis)aligned with the CMB dipole direction. The statistical significance in QSOs alone is $\gtrsim 2 \sigma$ and when combined with similar trends in strong lensing, Type Ia SN and calibrated GRBs, this increases to $\sim 3 \sigma$. Our findings are consistent with reported discrepancies in the cosmic dipole and anisotropies in galaxy cluster scaling relations. The reported variations in $H_0$ across the sky suggest that Hubble tension may be a symptom of a deeper cosmological malaise. 
\end{abstract}

\maketitle

\section{Introduction} 
Persistent cosmological tensions \cite{Riess:2020fzl, Freedman:2021ahq, Pesce:2020xfe, Blakeslee:2021rqi, Kourkchi:2020iyz, DES:2021wwk, KiDS:2020suj} suggest that
it is timely to reflect on the success of the flat $\Lambda$CDM cosmology based on Planck values \cite{Planck:2018vyg}. In particular, a $\sim 10\%$ discrepancy in the scale of the Hubble parameter in the post Planck era, \textit{if true},  belies the moniker ``precision cosmology". Recently, the community has gone to considerable efforts to address %make sense of 
these discrepancies (see \cite{DiValentino:2021izs}), but proposals are often physically contrived.  Great progress has been made in cosmology through the \textit{assumption} that the Universe is isotropic and homogeneous, namely the Cosmological Principle or Friedmann-Lema\^itre-Robertson-Walker (FLRW) paradigm. Nevertheless, cosmological tensions point to something being amiss. Here, building on an earlier companion letter \cite{Krishnan:2021jmh}, we present further evidence that FLRW may be the origin of cosmological tensions \cite{Krishnan:2021dyb} (see \cite{Fleury:2013uqa} for earlier comments). 

The Cosmic Microwave Background (CMB) dipole is almost ubiquitously assumed to be of kinematical origin, i. e. due to relative motion. By subtracting the dipole, the CMB is defined as the rest frame for the Universe. CMB anomalies have been documented in \cite{Schwarz:2015cma} and some refer to anomalies with directional dependence, for example the (planar) alignment of the quadrupole and octopole and their normals with the CMB dipole \cite{deOliveira-Costa:2003utu, Schwarz:2004gk}. In addition, it has been argued that an anomalous parity asymmetry \cite{Land:2005jq} may be traced to the CMB dipole \cite{Kim:2010gf, Naselsky:2011jp}, so a common origin for CMB anomalies is plausible.  

Separately, attempts to recover the CMB dipole from counts of late Universe sources such as radio galaxies \cite{Blake:2002gx, Singal:2011dy, Gibelyou:2012ri, Rubart:2013tx, Tiwari:2015tba, Colin:2017juj, Bengaly:2017slg, Siewert:2020krp} and QSOs \cite{Secrest:2020has}, which are assumed to be in ``CMB frame", largely agree that the CMB dipole direction is recovered, \textit{but not the magnitude}. The implication is that observables in the late Universe are not in the same FLRW Universe. Independently, similar findings have emerged from studies of the apparent magnitudes of Type Ia supernovae (SN) \cite{Singal:2021crs} and QSOs \cite{Singal:2021kuu}.  In contrast, internal analysis of CMB data confirms the CMB dipole magnitude, but with large errors that allow an intrinsic (non-kinematic) component \cite{Ferreira:2020aqa,Saha:2021bay}.\footnote{Interesting spatial dependence in the fine structure constant \cite{Webb:2010hc, King:2012id} and alignments in QSO polarisations \cite{Hutsemekers:2000fv, Hutsemekers:2005iz} have been reported elsewhere. The latter define an axis closely aligned with the CMB dipole.} It should be stressed that 
%although the statistics may be impressive,
these results are based on partial sky coverage and this is a potential systematic \footnote{It has been suggested that increased sky coverage helps with the lensing anomaly in the CMB \cite{Efstathiou:2019mdh}.}. 

Without doubt, the bread and butter of FLRW cosmology is the Hubble parameter $H(z)$. In particular, Hubble tension \cite{Riess:2020fzl, Freedman:2021ahq, Pesce:2020xfe, Blakeslee:2021rqi, Kourkchi:2020iyz} casts a spotlight on $H_0=H(z=0)$. Here, we build on earlier observations for strongly lensed QSOs \cite{Krishnan:2021dyb} and Type Ia SN \cite{Krishnan:2021jmh} that $H_0$ values in the direction of the CMB dipole, loosely defined, are larger with significance $1.7 \sigma$. Similar variations of $H_0$ across the sky have been reported for scaling relations in galaxy clusters \cite{Migkas:2020fza, Migkas:2021zdo}.\footnote{Curiously, \cite{Migkas:2020fza, Migkas:2021zdo} find that $H_0$ is lower along the CMB dipole direction, but restrict $\Omega_{m}$ to a Planck value in the analysis. This difference is worth investigating.} Note, within FLRW the value of $H_0$ is insensitive to the number of observables in any given direction, but of course the number of observables impacts the errors. For this reason, once one works with the Hubble diagram, sky coverage is not an expected systematic. Finally, a variation in $H_0$ across the sky may recast the Hubble tension discussion \cite{Riess:2020fzl, Freedman:2021ahq, Pesce:2020xfe, Blakeslee:2021rqi, Kourkchi:2020iyz} as a symptom of a deeper issue.  

Our findings are that QSOs, on the assumption that they represent {\textit{standardisable candles}} \cite{Risaliti:2015zla}, return higher $H_0$ values in hemispheres aligned with the CMB dipole direction at a significance of $ \gtrsim 2 \sigma$. See also \cite{Zhao:2021zih} for overlapping analysis. However, similar analysis in \textit{standardisable} GRBs \cite{Schaefer:2007s, Wang:2007, Amati:2008hq, Capozziello:2008, Dainotti:2008, Izzo:2009, Amati:2013, Wei:2014, Izzo:2015, Tang:2019} fails to show the same trend and is consistent with an isotropic Hubble expansion. Relative to the QSOs, it is worth stressing that GRB samples are much smaller and the intrinsic scatter is larger, so the difference in results may not be so surprising. Nevertheless, noting that the QSOs are distributed anisotropically on the sky in a manner that is  (mis)aligned with the CMB dipole direction, we trim the GRB samples by removing GRBs less closely (mis)aligned with the CMB dipole direction. For uncalibrated GRBs, we find no effect, whereas for GRBs calibrated by Type Ia SN, larger values of $H_0$ in the CMB dipole direction are found at up to $1.7 \sigma$.

Admittedly, in contrast to Type Ia SN, QSOs and GRBs are non-standard, but \textit{if} they are \textit{standardisable}, then they should be able to track $H_0$, namely a universal constant in \textit{all} FLRW cosmologies (see discussion in \cite{Krishnan:2020vaf}). Objectively, QSOs and GRBs constitute emerging cosmological probes \cite{Moresco:2022phi}. We emphasise that the physics of strong lensing time delay, Type Ia SN, QSOs and GRBs are sufficiently different with varying systematics. We quantify the significance of these observations by adopting different combinations of data, different redshift ranges and different orientations (that are expected to be irrelevant in a strict FLRW sense) arriving at significances in the range $1.7 \sigma - 3 \sigma$. Although the significance is below the discovery threshold, it should be borne in mind that we are discussing a universal parameter that \textit{must} be a constant within FLRW cosmology. If there is any chance that it is not a constant, the potential consequences are far-reaching. The lower $1.7 \sigma$ bound is conservative and rests solely on strong lensing \cite{Wong:2019kwg, Millon:2019slk} and Pantheon Type Ia SN \cite{Scolnic:2017caz}. Once QSOs \cite{Lusso:2020pdb} are taken into account, the significance is in the $2.4-2.7\sigma$ window, while GRBs \cite{Demianski:2016zxi} potentially push this to $3 \sigma$. It is hence plausible that at cosmic distances in the redshift range $O(0.1) \lesssim z\lesssim O(1)$ (and possibly beyond) the Hubble expansion is anisotropic and we have a preferred  direction along the CMB dipole.\footnote{Note, it is already an established fact that at lower redshifts, in particular distances out to 200 Mpc, the local Universe is anisotropic, e. g. \cite{Hoffman:2017ako}.} As explained in \cite{Krishnan:2021dyb}, future observations of strongly lensed QSOs  \cite{Wong:2019kwg, Millon:2019slk} and potentially lensed SN \cite{Suyu:2020opl} may settle the issue.

\section{QSO DATA}
\label{sec:QSO}
QSOs as {{standardisable candles}} in cosmology would be a game changer, since they are plentiful even up to redshift  $z \sim 7$. Despite a number of competitive proposals, e. g. \cite{Watson:2011um, Wang:2013ha, LaFranca:2014eba} (see also \cite{Dai:2012wp, Solomon:2021jml}), arguably the simplest and most powerful approach, due to Risaliti \& Lusso \cite{Risaliti:2015zla}, exploits an empirical relation between X-ray and UV luminosities in QSOs \cite{Tananbaum}, 
\be
\label{X_UV}
\log_{10} L_{X} = \beta + \gamma \log_{10} L_{UV}, 
\ee
where $\beta $ and $\gamma \approx 0.6$ are constants. Various studies have shown the robustness of the slope $\gamma$ both over orders of magnitude in luminosity and over extended redshift ranges \cite{Vignali:2002ct, Just:2007se, Lusso:2009nq, Salvestrini:2019thn, Bisogni:2021hue}. The program \cite{Risaliti:2015zla, Risaliti:2018reu, Lusso:2019akb, Lusso:2020pdb} is still in its infancy and reminiscent of the status of Type Ia SN in the 1990s \cite{Phillips:1993ng}. As we touch upon later, some corrections for running in $(\beta, \gamma)$ may still be required, but this is an ongoing and active research line. Of course, the same is true for standardisable SN. The results so far have been intriguing, especially since they are at odds with Planck-$\Lambda$CDM \cite{Planck:2018vyg}. 

In particular, QSOs (and GRBs) prefer larger values of matter density $\Omega_{m}$, consistent with a Universe with little or no dark energy \cite{Yang:2019vgk, Velten:2019vwo, Khadka:2020whe, Khadka:2020vlh} \footnote{As explained in \cite{Banerjee:2020bjq}, the results of \cite{Risaliti:2018reu, Lusso:2019akb, Lusso:2020pdb} have been negatively impacted by the cosmographic expansion, so claims of discrepancies in $\Omega_{m}$ cannot be substantiated without outside analysis.}. As is clear from Fig. 6 of \cite{Horstmann:2021jjg} (based on \cite{Dainotti:2021pqg}), the same trend is also there in HST SN. Evidently, this is due to a preference within QSOs, GRBs and HST SN for lower luminosity distances $D_{L}(z)$ relative to Planck-$\Lambda$CDM \cite{Planck:2018vyg}, especially at higher redshifts. Interestingly, Lyman-$\alpha$ BAO \cite{BOSS:2014hwf, duMasdesBourboux:2020pck} also prefers lower values of angular diameter distance $D_{A}(z) = (1+z)^{-2} D_{L}(z)$ and is therefore consistent with the Risaliti-Lusso QSOs \footnote{The DES collaboration has also recently reported lower values of $D_{A}(z)$ at an effective redshift of $z_{\textrm{eff}} = 0.835$ through a blind analysis \cite{DES:2021esc}.}.  
Separately, it has been argued that best fit values of $\beta, \gamma$ may be sensitive to the cosmological model, suggesting they should only be employed over restricted redshift ranges, $z \lesssim 1.7$ \cite{Khadka:2020tlm}, or with QSO subsamples \cite{Khadka:2021xcc}. We address this concern in section \ref{sec:qso_caveats}, while in appendix \ref{sec:QSOHD} we comment more generally on the status of the QSO Hubble diagram and explain how some remnant evolution in ($\beta, \gamma$) with redshift is unlikely to mimic our directional results. 

\subsection{Methods} 
The key idea of Risaliti \& Lusso  \cite{Risaliti:2015zla} is to assume that the relation (\ref{X_UV}) holds, before converting it into a relation in UV and X-ray fluxes,  $F_{UV}$ and $F_{X}$, respectively:
\be
\label{fluxes}
\log_{10} F_{X} = \beta  + \gamma \log_{10} F_{UV} +  (\gamma-1)  \log_{10} (4\pi D_{L}^2).
%\log_{10} F_{X} = \beta + (\gamma-1) \log_{10}  4 \pi + \gamma \log_{10} F_{UV} + 2 (\gamma-1)  \log_{10} (D_{L}).
\ee
In a flat cosmology, $D_{L}(z)$ may be expressed as 
\be
D_{L} (z) = c (1+z) \int_{0}^{z} \frac{1}{H(z')} \textrm{d} z'. 
\ee
Owing to the scatter in the QSO data, an additional intrinsic dispersion parameter $\delta$ is considered \cite{Risaliti:2015zla}. Thus, within the flat $\Lambda$CDM model, in addition to two cosmological parameters, ($H_0, \Omega_{m}$), there are three extra parameters: $\beta, \gamma$ from (\ref{fluxes}) and $\delta$. However, $\beta$ is degenerate with $\log_{10}H_0$, so both cannot be independently determined without external data. Here we do not combine data sets, as the poorer quality QSO data risks returning unrepresentative values \cite{Yang:2019vgk} and the goal is to extract information from QSOs directly. Due to this degeneracy we may fix $H_0$ to a canonical value, e.g. $H_0=70$ km/s/Mpc. In our analysis here we mainly focus on the variation of a quantity over putative hemispheres in the sky, which for observable $X$ will be denoted by $\Delta X$.  From \eqref{fluxes} one observes that $\Delta \beta\propto \Delta H_0/H_0$.  
%Therefore, we fix $H_0$ to the canonical value $H_0 = 70$ km/s/Mpc and fit $\beta$ as a proxy for $H_0$. 
Since $\gamma \approx 0.6 < 1$, the proportionality factor is positive, i.e. an \textit{increase} in $\beta$ is equivalent to an \textit{increase} in $H_0$. 

The best-fit parameters $(\Omega_{m}, \beta, \gamma, \delta)$ follow from extremising the likelihood function \cite{Risaliti:2015zla}, 
\be
\label{L1}
\mathcal{L} = - \frac{1}{2} \sum_{i=1}^{N} \left[ \frac{ \left(\log_{10} F^{\textrm{obs}}_{X, i} - \log_{10} F^{\textrm{model}}_{X,i}\right)^2}{s^2_{i}} + \ln (2 \pi s_i^2) \right], 
\ee 
where $s_i^2 = \sigma_i^2+ \delta^2$ contains the measurement error on the observed flux $\log_{10} F^{\textrm{obs}}_{X,i}$ and $\delta$. $\log_{10} F^{\textrm{model}}_{X,i}$ carries information about the cosmological model through (\ref{fluxes}).

\begin{figure}[htb]
\centering
  \includegraphics[width=80mm]{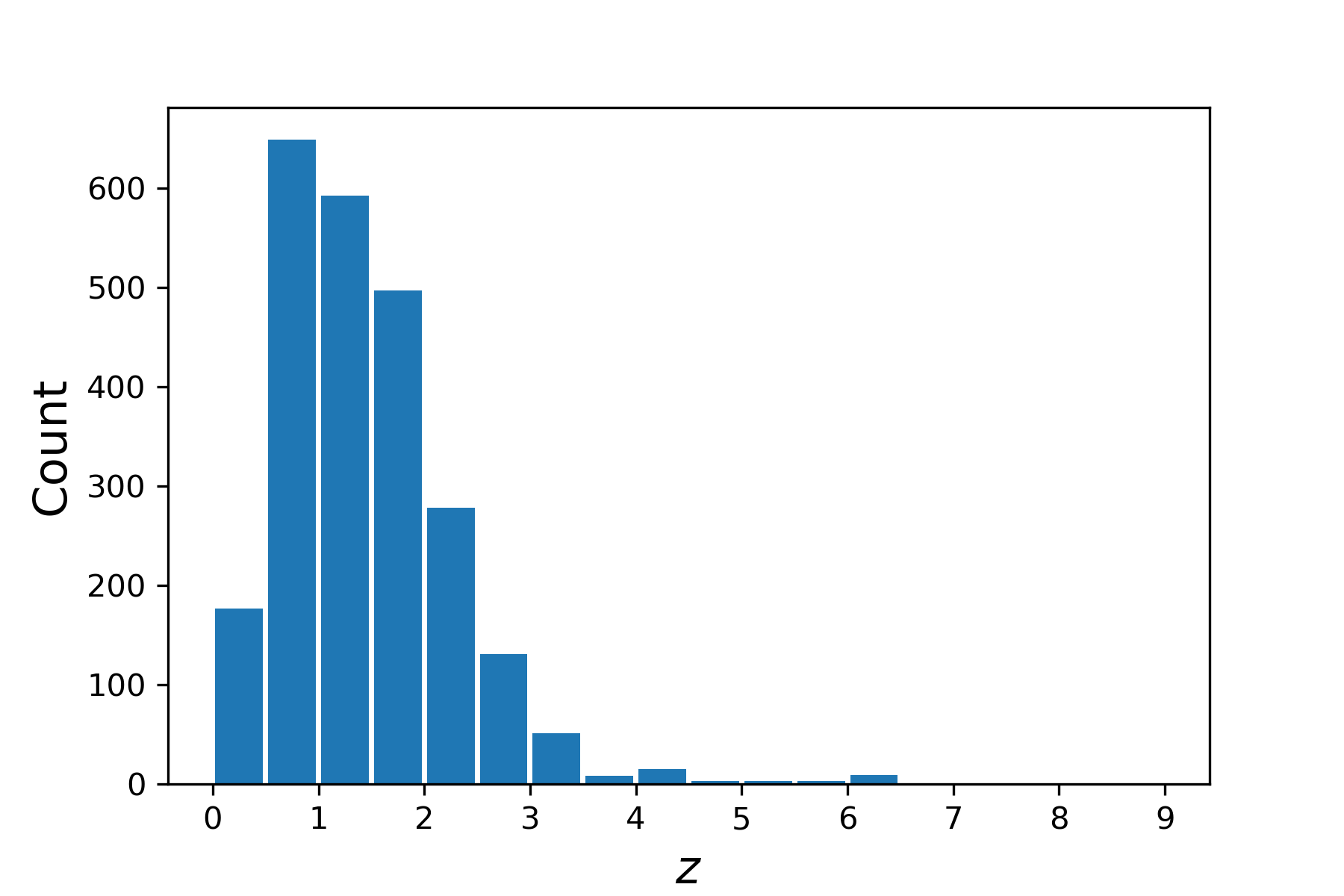}
\caption{The redshift distribution of the 2421 QSOs from the recent compilation \cite{Lusso:2020pdb} in intervals of $\Delta z = 0.5$.}
\label{qso_count}
\end{figure}

\begin{figure}[htb]
\centering
  \includegraphics[width=70mm]{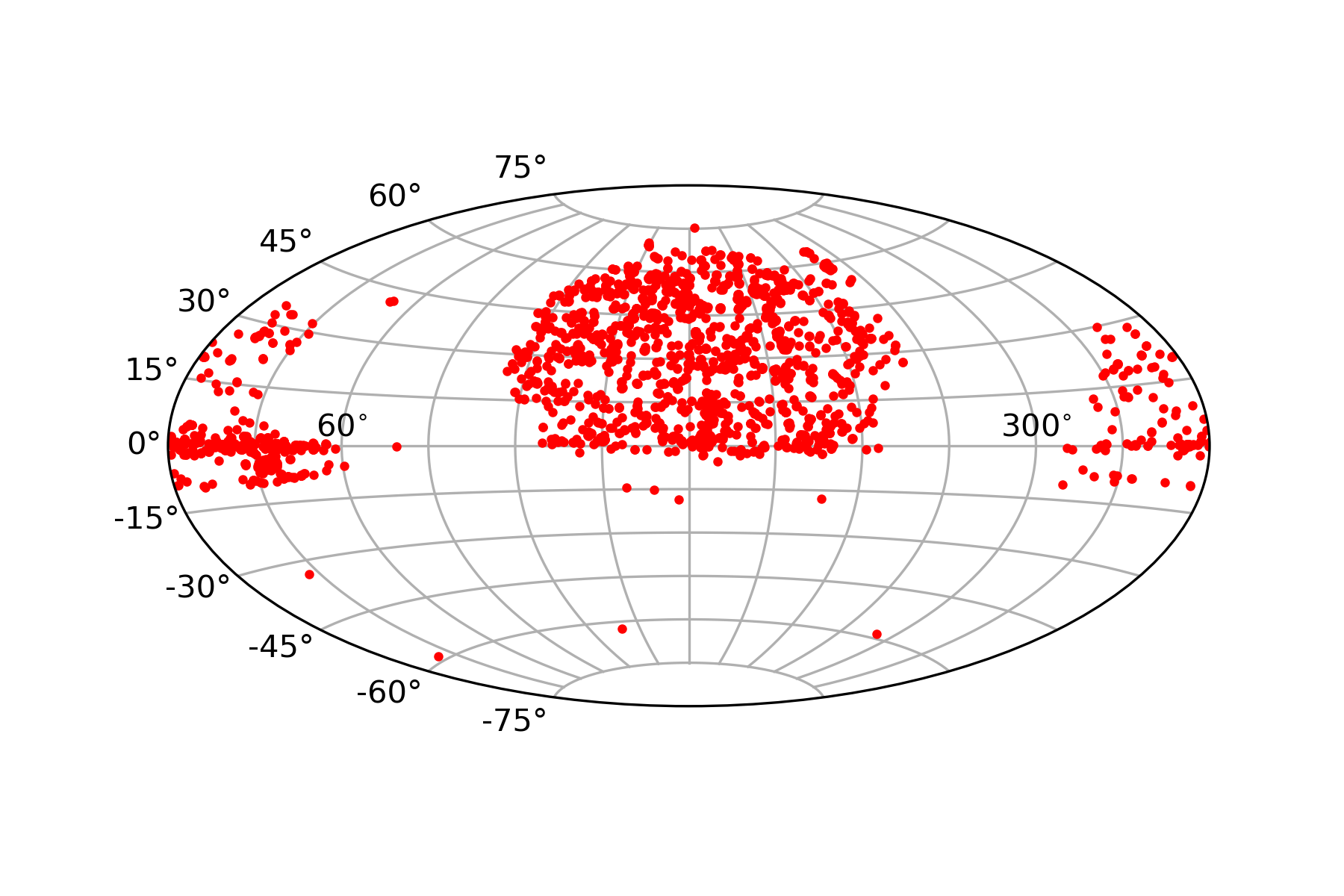}
\caption{Distribution of the QSOs  \cite{Lusso:2020pdb} on the sky.}
\label{qso_loc}
\end{figure}

On the data side, we make use of the latest compilation of QSO data \cite{Lusso:2020pdb}, which contains 2421 QSOs in the redshift range $0.009 \leq z \leq 7.5413$. We show the redshift distribution of the full QSO sample and its distribution on the sky in  FIG. \ref{qso_count} and FIG. \ref{qso_loc}. Evidently, the data becomes sparse as one approaches $z=4$, while it is noticeable that the majority of the QSOs, 1655 in fact, are located in the range $ 90^{\circ} < \textrm{RA} < 270^{\circ}$ and in the northern hemisphere, $ \textrm{DEC} > 0^{\circ}$. Here, RA and DEC denote right ascension and declination. This means that the QSO distribution on the sky is anisotropic, but working within the FLRW assumption, this is not expected to make any difference. We will return to study anisotropic distributions later with GRBs. As explained in Ref. \cite{Lusso:2020pdb}, while one can use the overall data set, the UV fluxes for some $z < 0.7$ QSOs have been determined by extrapolation from the optical, however there are some local QSOs ($ z < 0.1$) whose UV spectra have been determined without extrapolation and one can have greater confidence in them. While one can include the local QSOs, and we do in appendix \ref{sec:QSOHD} to confirm that QSOs recover the same Planck-$\Lambda$CDM Universe where they overlap well with SN, it is conservative to remove all the QSOs below $z = 0.7$ \cite{Lusso:2020pdb} and this reduces the sample to 2023 QSOs. Observe that $z=0.7$ is large enough that all the QSOs are expected to be in the same FLRW Universe as the CMB. Peculiar velocities are not relevant.  

\subsection{Analysis}
We start by performing a consistency check in a bid to recover results quoted in \cite{Khadka:2020tlm}. To that end, we retain the local QSOs ($z < 0.1$), which we combine with QSOs in the redshift range $0.7 < z < 1.479$.\footnote{\label{footie} In the data set downloaded from VizieR, we are unable to find the last two entries in Table 2 of \cite{Lusso:2020pdb}, otherwise we are using the same local QSOs.} Throughout we use the flat priors $0 \leq \Omega_{m} \leq 1$, $0 < \beta < 15$, $0 < \gamma < 1$ and $ 0 < \delta < 1$. As is clear from Table \ref{warmup}, the results of extremisation and marginalisation via MCMC agree well, modulo the fact that $\Omega_{m}$ is displaced to smaller values. We have checked that the best-fit $\Omega_{m}$ value corresponds to the peak of the distribution, at least within the bounds, which means once restricted to the range $0 \leq \Omega_{m} \leq 1$, the distribution is lopsided, so the marginalised values are shifted to smaller values. In effect, $\Omega_{m}$ wants to exceed the bound in order to reduce $D_{L}(z)$, but within flat $\Lambda$CDM, it cannot. Thus, displacements in MCMC values of $\Omega_{m}$ are an artefact of the bounds, otherwise extremisation and MCMC show good agreement. Finally, we can compare the results from \cite{Khadka:2020tlm}, reproduced in Table \ref{warmup} and confirm that there is  agreement, despite a slight difference in data (see footnote \ref{footie}). 

In appendix \ref{sec:QSOHD} we show that $\Omega_{m}$ gradually increases from $\Omega_{m} \approx 0.3$ to $\Omega_{m} \approx 1$ as redshift ranges are extended beyond the traditional SN range. In short, QSO data are weighted to much higher redshifts. This challenges preconceptions based on SN cosmology, which are biased to lower redshifts and thus well anchored in the dark energy dominated regime. In contrast, the QSO distribution is such that the relatively small number of anchoring low redshift QSOs are not statistically significant enough to prevent QSOs seeing an Einstein-de Sitter Universe composed simply of matter. As we explain in appendix \ref{sec:QSOHD}, one gets similar results in SN provided one removes the anchoring low redshift SN in sufficient number. Therefore, while $\Omega_{m} \approx 1$ may look strange, it is arguably the correct result for high redshift probes that are not sufficiently anchored in the late Universe.

\begin{table}[htb]
\centering 
\begin{tabular}{c|c|c|c}
 
\rule{0pt}{3ex} $ \Omega_{m}$ & $ \beta$ & $\gamma$ & $\delta$ \\
\hline 
\rule{0pt}{3ex} $0.843$ & $9.110$ & $0.589$ & $0.238$  \\
\rule{0pt}{3ex} $0.697^{+0.204}_{-0.238}$ & $8.980^{+0.505}_{-0.531}$ & $0.594^{+0.017}_{-0.016}$ & $0.239^{+0.006}_{-0.006}$  \\
\hline 
\rule{0pt}{3ex} $0.800$ & $8.695$ & $0.584$ & $0.238$  \\
\rule{0pt}{3ex} $0.670^{+0.300}_{-0.130}$ & $8.570^{+0.530}_{-0.530}$ & $0.588^{+0.018}_{-0.018}$ & $0.239^{+0.006}_{-0.006}$  \\

\end{tabular}
\caption{The best-fit and marginalised values of the parameters for QSOs in the redshift range $0 < z < 0.1 \cup 0.7 < z < 1.479$. The first line corresponds to extremising the likelihood (\ref{L1}), whereas the second follows from an MCMC exploration, where we quote $1 \sigma$ confidence intervals. The third and fourth lines record the analogous results from \cite{Khadka:2020tlm}, modulo a slight difference in data (footnote \ref{footie}).}
\label{warmup}
\end{table}

The take away from the warmup exercise is that both extremisation and marginalisation via MCMC return consistent values of $\beta$. For us this is important, as we will explore variations of  $\beta$ across the sky by scanning over  RA, DEC and using extremisation to identify differences in absolute $\beta$ values between hemispheres. Extremisation with a focus on $\beta$ is  considerably quicker than MCMC, or fitting the logarithm dressed $H_0$, and once the variations in $\beta$ have been identified, we drill down on the more interesting orientations using MCMC in order to quantify the errors and extract the significance of any discrepancy. Concretely, we break the sky up into a $31 \times 15$ grid. Each point on this grid corresponds to two angles, which can be traded for a vector \cite{Krishnan:2021jmh}, 
\be
\vec{v} = [ \cos (\textrm{DEC}) \cos (\textrm{RA}), \cos (\textrm{DEC}) \sin (\textrm{RA}), \sin (\textrm{DEC}) ].  
\ee
Observe that one gets the antipodal point on the sky by flipping the sign of \textrm{DEC} and shifting \textrm{RA} by $180^{\circ}$, so by opting for an odd number of points in our grid, we include antipodal points. This duplication allows for consistency checks. Next, one separates the sample based on the sign of the inner product of this vector with the corresponding vector for each data point in the QSO sample. This splits the data into two hemispheres. Once done, one extremises the likelihood (\ref{L1}) for each hemisphere and records the difference between the ``northern" (N) and ``southern" (S) hemisphere, $\Delta \beta = \beta^{N} - \beta^{S}$. 

\begin{figure}[htb]
\centering
  \includegraphics[width=80mm]{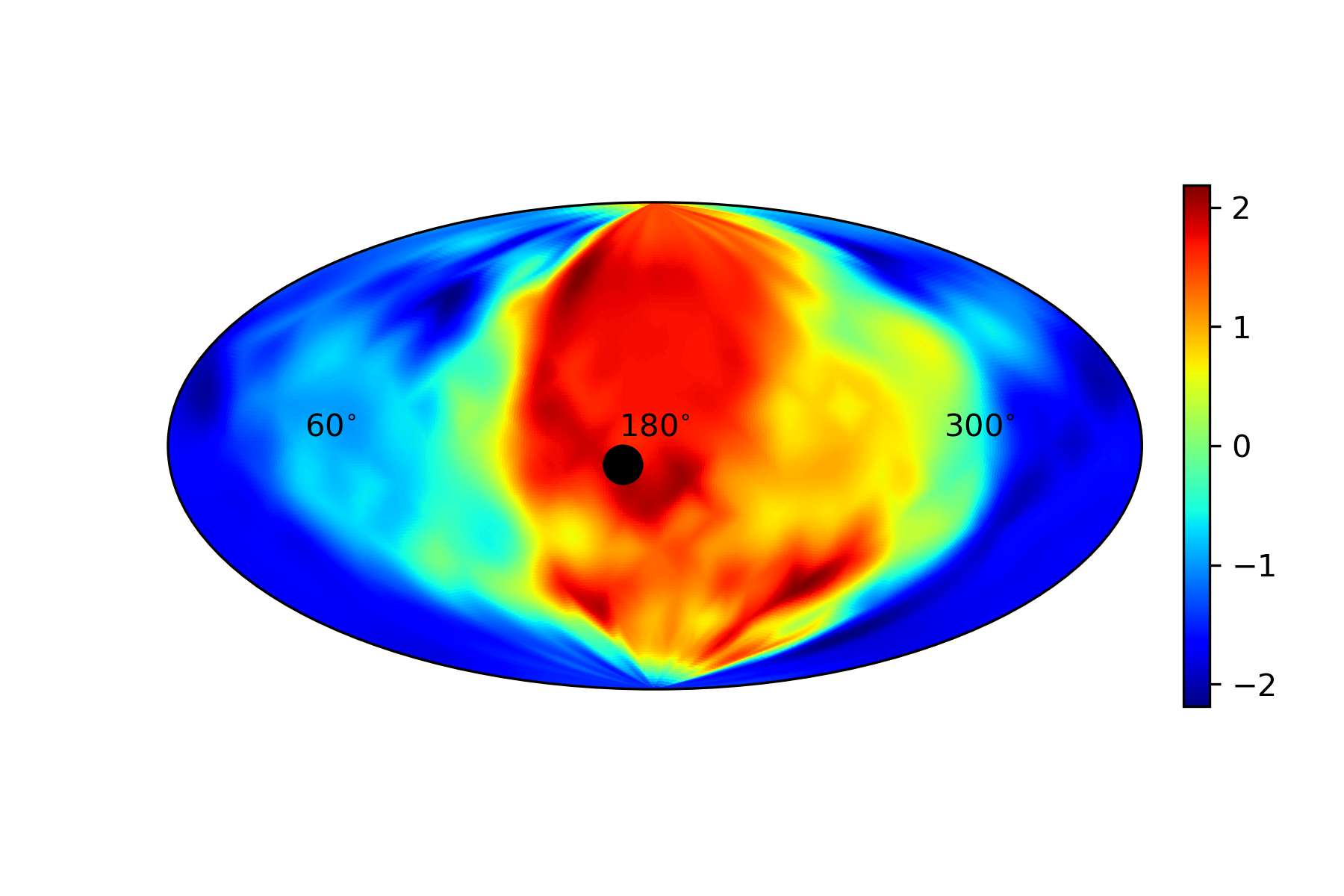}
\caption{Variations \green{of} the best-fit $\beta$ parameters in respective hemispheres as (RA, DEC) values for the QSOs in the redshift range $0.7 < z \leq 7.5413$. The black dot denotes the CMB dipole. The lower (higher) $\beta$ region corresponds to lower (higher) $H_0$ regions.}
\label{qso_dipole}
\end{figure}

The result of this scan over the angles is shown in FIG. \ref{qso_dipole}, where we have included the CMB dipole direction $(\textrm{RA}, \textrm{DEC}) = (168^{\circ}, -7^{\circ})$ for guidance, and used the python library \textit{scipy} (scipy.interpolate.griddata) to perform a cubic interpolation in $\Delta \beta$. We have checked that the antipodal point on the sky simply flips the sign of $\Delta \beta$.  For this reason, the mean (and median) of our distribution in $\Delta \beta$ coincides with $\Delta \beta = 0$ and we have confirmed this is the case. In Table \ref{qso} we record the best-fit and marginalised parameters for the CMB dipole direction and the direction of maximum $\Delta \beta$, where we have suppressed $\delta$ as it shows little variation. Note, we only consider the maximum $\Delta \beta$ from the sampled points and not the interpolation. 

\begin{table}[htb]
\centering 
\begin{tabular}{c|c|c|c|c}
\rule{0pt}{3ex} (RA, DEC)& Hemisphere & $ \Omega_{m}$ & $ \beta$ & $\gamma$ \\
\hline 
\rule{0pt}{3ex} \multirow{ 4}{*}{CMB dipole} &  \multirow{ 2}{*}{N} & $1$ & $8.491$ & $0.609$  \\
\rule{0pt}{3ex} & & $0.924^{+0.057}_{-0.107}$ & $8.451^{+0.356}_{-0.371}$ & $0.610^{+0.012}_{-0.011}$ \\
\cline{3-5}
\rule{0pt}{3ex}  & \multirow{ 2}{*}{S} & $1$ & $6.787$ & $0.663$  \\
\rule{0pt}{3ex} & & $0.880^{+0.087}_{-0.155}$ & $6.714^{+0.535}_{-0.538}$ & $0.666^{+0.017}_{-0.017}$ \\
\hline 
\rule{0pt}{3ex} \multirow{ 4}{*}{$(132^{\circ}, 64.3^{\circ})$} & \multirow{ 2}{*}{N} & $1$ & $8.474$ & $0.609$ \\
\rule{0pt}{3ex} & & $0.934^{+0.048}_{-0.096}$ & $8.426^{0.354}_{-0.343}$ & $0.611^{+0.011}_{-0.011}$\\
\cline{3-5}
\rule{0pt}{3ex}  & \multirow{2}{*}{S} & $1$ & $6.291$ & $0.679$ \\ 
\rule{0pt}{3ex}  & & $0.845^{+0.114}_{-0.191}$ & $6.171^{+0.633}_{-0.594}$ & $0.683^{+0.019}_{-0.020}$ \\
\end{tabular}
\caption{Best-fit values of the parameters from both extremisation and MCMC marginalisation of the likelihood (\ref{L1}) for QSOs in the redshift range $0.7 < z \leq 7.5413$.}
\label{qso}
\end{table}

As is clear from Table \ref{qso}, once again extremisation and MCMC show good agreement. We have randomly sampled other points to confirm that this agreement is more widespread. From the $1 \sigma$ confidence intervals in Table \ref{qso} one can estimate the discrepancy in $\Delta \beta$ between hemispheres to be  $2.7 \sigma$ for the CMB dipole and  $3 \sigma $ for the maximum $\Delta \beta$. It is easy to check that both directions are in the same hemisphere using the vector inner product. These results may not be so surprising since we are working with the same tracer (QSOs) where a mismatch in the cosmic dipole has been reported with the CMB dipole at $\sim 5 \sigma$ \cite{Secrest:2020has}. From our end, FIG. \ref{qso_dipole} is reminiscent of similar features in strong lensing time delay \cite{Krishnan:2021dyb} and Type Ia SN \cite{Krishnan:2021jmh}. 

\begin{figure}[htb]
\centering
  \includegraphics[width=80mm]{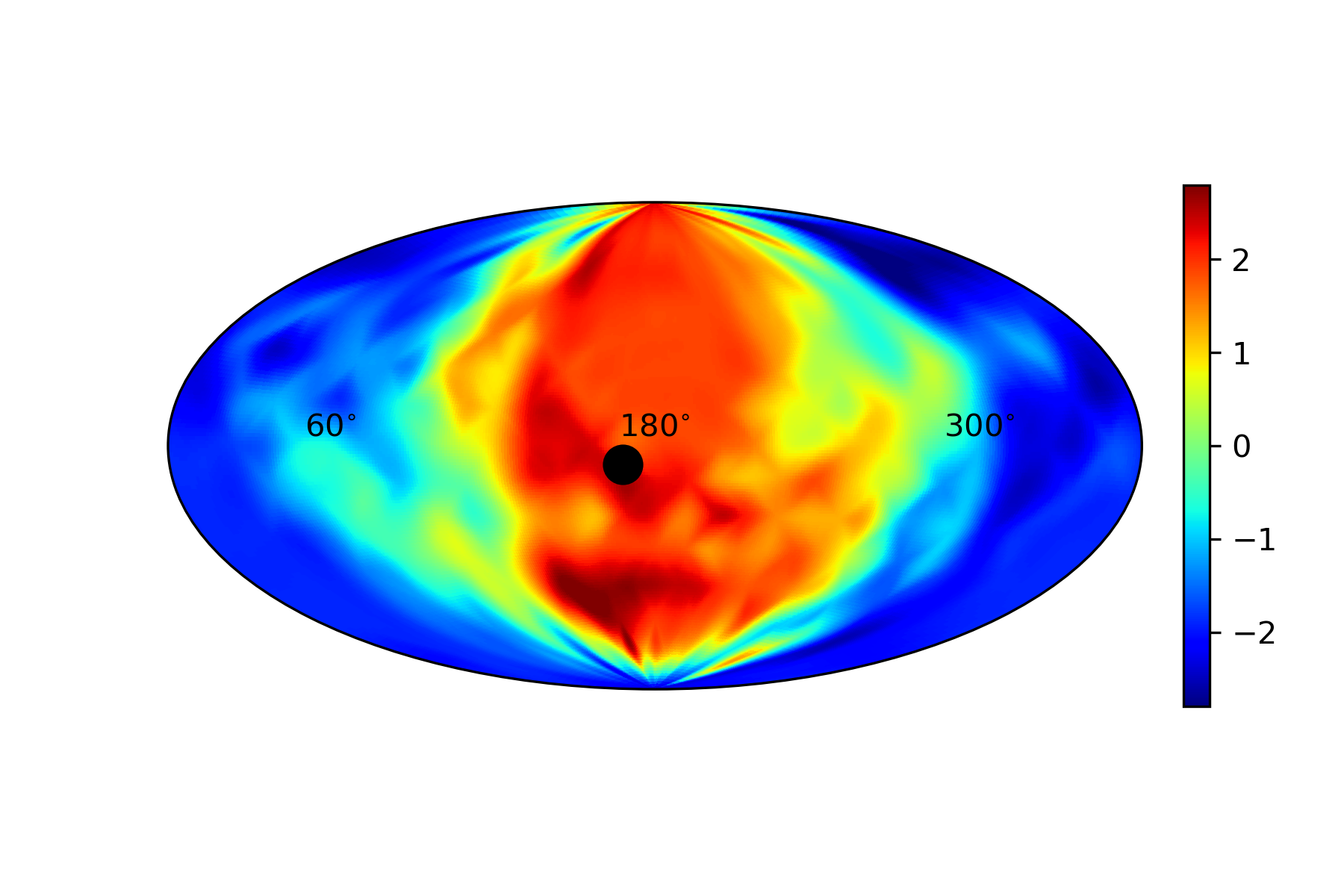}
\caption{Variations \green{of} the best-fit $\beta$ parameters in respective hemispheres as (RA, DEC) values for the QSOs in the redshift range $0.7 < z < 1.7$. The black dot denotes the CMB dipole.}
\label{qso_dipole_z17}
\end{figure}

We address caveats later in section  \ref{sec:qso_caveats} where we comment on concerns that $\beta$ evolves with redshift and varies across cosmological models \cite{Khadka:2020tlm}.
However, here it is prudent to repeat our analysis in the more conservative redshift range $ 0.7 < z < 1.7$.  The corresponding plot and results can be found in FIG. \ref{qso_dipole_z17} and Table \ref{qso_z17}. As expected, with the restricted redshift range, we reduce the QSO count to 1255, so this inflates the errors and reduces the significance. That being said, we still find a $2 \sigma$ discrepancy for the CMB dipole and $2. 8 \sigma$ for the maximum $\Delta \beta$. Once again, these directions are in the same hemisphere as the CMB dipole. Interestingly, the maximum $\Delta \beta$ direction has flipped hemisphere, but this may be expected: FLRW $H_0$ is an extremely blunt probe of any anisotropy. Nevertheless, qualitatively FIG. \ref{qso_dipole_z17} is the same as FIG. \ref{qso_dipole}. 

\begin{table}[htb]
\centering 
\begin{tabular}{c|c|c|c|c}
\rule{0pt}{3ex} (RA, DEC)& Hemisphere & $ \Omega_{m}$ & $ \beta$ & $\gamma$ \\
\hline 
\rule{0pt}{3ex} \multirow{ 4}{*}{CMB dipole} &  \multirow{ 2}{*}{N} & $1$ & $9.527$ & $0.575$  \\
\rule{0pt}{3ex} & & $0.694^{+0.211}_{-0.261}$ & $9.435^{+0.589}_{-0.573}$ & $0.579^{+0.018}_{-0.019}$ \\
\cline{3-5}
\rule{0pt}{3ex}  & \multirow{ 2}{*}{S} & $1$ & $7.604$ & $0.636$  \\
\rule{0pt}{3ex} & &  $0.762^{+0.168}_{-0.273}$ & $7.489^{+0.805}_{-0.788}$ & $0.641^{+0.025}_{-0.026}$ \\
\hline 
\rule{0pt}{3ex} \multirow{ 4}{*}{$(132^{\circ}, -51.4^{\circ})$} & \multirow{ 2}{*}{N} & $0.678$ & $11.199$ & $0.524$ \\
\rule{0pt}{3ex} & & $0.587^{+0.283}_{-0.294}$ & $11.084^{+0.898}_{-0.917}$ & $0.528^{+0.030}_{-0.028}$ \\
\cline{3-5}
\rule{0pt}{3ex}  & \multirow{2}{*}{S} & $1$ & $8.206$ & $0.617$ \\ 
\rule{0pt}{3ex}  & & $0.806^{+0.139}_{-0.214}$ & $8.121^{+0.537}_{-0.563}$ & $0.620^{+0.018}_{-0.017}$  \\
\end{tabular}
\caption{Best-fit values of the parameters from both extremisation and MCMC marginalisation of the likelihood (\ref{L1}) for QSOs in the redshift range $0.7 < z < 1.7$.}
\label{qso_z17}
\end{table}

\subsection{Simulations} 
Taken at face value, FIG. \ref{qso_dipole} and FIG. \ref{qso_dipole_z17} reveal striking emergent dipoles. As we have shown, one can find orientations in the sky where the variations  in $\beta$,  $\Delta \beta$, is significant. Nevertheless, this is an \textit{a posteriori} inference. Here, we will attempt to quantify the significance through simulations based on the \textit{a priori} assumption that the CMB dipole direction is relevant. Let us explain why this is the case. Recall that the CMB dipole has been subtracted on the assumption that it is purely kinematic in origin, i. e. due to relative motion. Existing results point to a persistent excess in the cosmic dipole with respect to the CMB value \cite{Blake:2002gx, Singal:2011dy, Gibelyou:2012ri, Rubart:2013tx, Tiwari:2015tba, Colin:2017juj, Bengaly:2017slg, Siewert:2020krp, Secrest:2020has, Singal:2021crs, Singal:2021kuu}, which if true, implies that CMB \cite{Planck:2018vyg}, radio galaxies \cite{Blake:2002gx, Singal:2011dy, Gibelyou:2012ri, Rubart:2013tx, Tiwari:2015tba, Colin:2017juj, Bengaly:2017slg, Siewert:2020krp}, QSOs \cite{Secrest:2020has, Singal:2021kuu} and Type Ia SN \cite{Singal:2021crs} \footnote{Recently, using corrected redshifts for Pantheon \cite{Steinhardt:2020kul}, Ref. \cite{Horstmann:2021jjg} reports a smaller magnitude of the cosmic dipole with respect to the CMB value.} do not inhabit the same FLRW frame. Our results are certainly consistent with this discrepancy. We are ultimately seeing that QSOs, which should be in the CMB rest frame, and thus ambivalent to the CMB dipole direction, are mysteriously tracking the CMB dipole direction.

It remains to see how unlikely this is in a representative flat $\Lambda$CDM Universe. To do so, we turn to simulations. Concretely, we first fix the redshift range, either $0.7 < z \leq 7.5413$ or $0.7 < z < 1.7$, and fit the flat $\Lambda$CDM model to the data to get an MCMC chain of representative values for $(\Omega_{m}, \beta, \gamma, \delta)$. As before, throughout we set $H_0 = 70$ km/s/Mpc. Next for each entry in the chain, we mock up new values of $\log_{10} F_{UV}$ by picking them randomly from normal distributions based on the corresponding error in $\log_{10} F_{UV}$. Here, the errors are typically small, so the new values are close to the original values of $\log_{10} F_{UV}$ and essentially all positive, $\log_{10} F_{UV} > 0$, as required. Next, we generate $\log_{10} F_{X}$ values consistent with equation (\ref{fluxes}) by employing $(\Omega_{m}, \beta, \gamma)$ from the MCMC chain to establish a mean value for $\log_{10} F_{X}$, before generating values in a normal distribution with standard deviation $s_i = \sqrt{\sigma_i^2 + \delta^2}$, where $\delta$ is also extracted from the MCMC chain. Doing this for entries in the MCMC chain, one builds up mock realisations of data that are consistent with both equation (\ref{fluxes}) and flat $\Lambda$CDM. We have checked that the mocking procedure typically returns values of ($\Omega_m, \beta, \gamma, \delta$) within the $1 \sigma$ confidence interval as defined by the MCMC chain.

\begin{figure}[htb]
\centering
  \includegraphics[width=80mm]{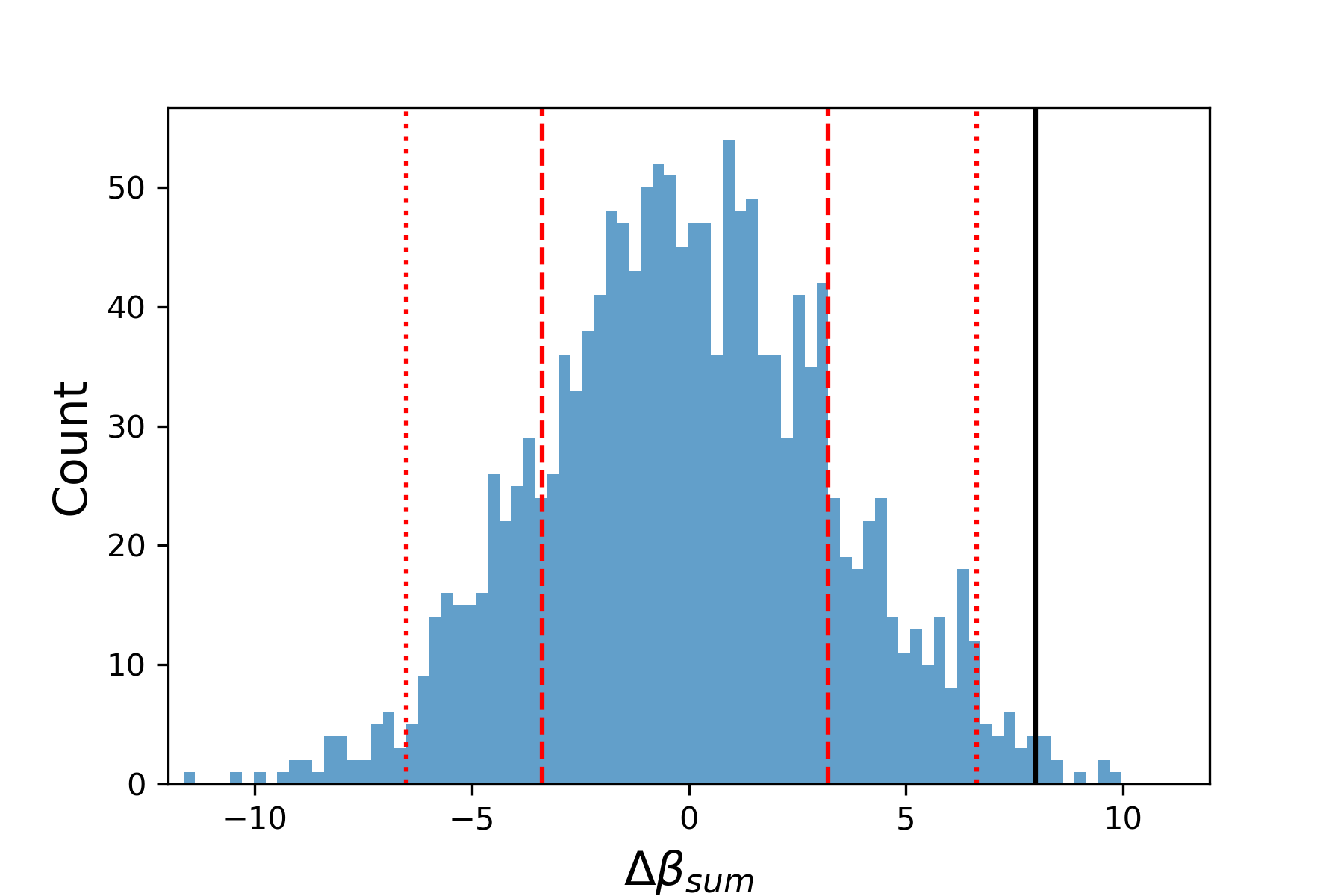}
\caption{The distribution of the weighted sum $\Delta \beta_{\textrm{sum}}$ for 1500 mock realisations of the real data from FIG. \ref{qso_dipole}. The red dashed and dotted lines represent $1 \sigma$ and $2 \sigma $, respectively, while the value for the real data, $\Delta \beta_{\textrm{sum}} = 7.98$, corresponds to the black line. }
\label{sim_qso}
\end{figure}

The question now is how often do figures such as FIG. \ref{qso_dipole} and FIG. \ref{qso_dipole_z17} arise as statistical fluctuations? To answer this, we condense the heat maps into a number. Now, this number could be simply the discrepancy in $\beta$ in a given direction, e. g. the CMB dipole direction, but it is better to focus on the hemisphere in the direction of the CMB dipole. The approach we adopt is to coarse grain our $31 \times 15$ grid to $11 \times 5$, which once overlapping points at the boundary of the RA range are removed, leaves us with 50 \textit{unassuming} points on the sky. From these 50, by symmetry 25 will be in the same hemisphere as the CMB dipole. Now, one could simply sum the discrepancies in $\beta$ at each point on the sky, i. e.  $\sum_{i=1}^{25} \Delta \beta_i$, but it is better to introduce a weighting so that points on the sky that are closest to the CMB dipole direction contribute the most. The reason being that evidence suggests the matter dipole is not exactly aligned with the CMB dipole, e. g. \cite{Siewert:2020krp, Secrest:2020has}. Therefore, our approach here is designed to allow some ambiguity in the direction of any putative matter dipole provided it is close to the CMB dipole. Concretely, the proposal is to use the weighted sum, 
\be
\label{weighted_sum} 
\Delta \beta_{\textrm{sum}} = \sum_{i=1}^{25} w_i\ \Delta \beta_i,  
\ee
where 
\be
\label{weight}
w_i :=\vec{v}_i \cdot \vec{v}_{\textrm{dipole}}, 
\ee 
and $\vec{v}_{\textrm{dipole}}$ denotes the vector in the direction of the CMB dipole. Note that we have added the subscript ``sum" to distinguish the weighted sum of $\Delta \beta_i$ and also that the inner product represents a natural way to weight the sum in the direction of the CMB dipole direction. This ensures dipoles that emerge as statistical fluctuations within the same hemisphere, \textit{yet in directions other than the CMB dipole direction}, make a less significant contribution to the weighted sum.

\begin{figure}[htb]
\centering
  \includegraphics[width=80mm]{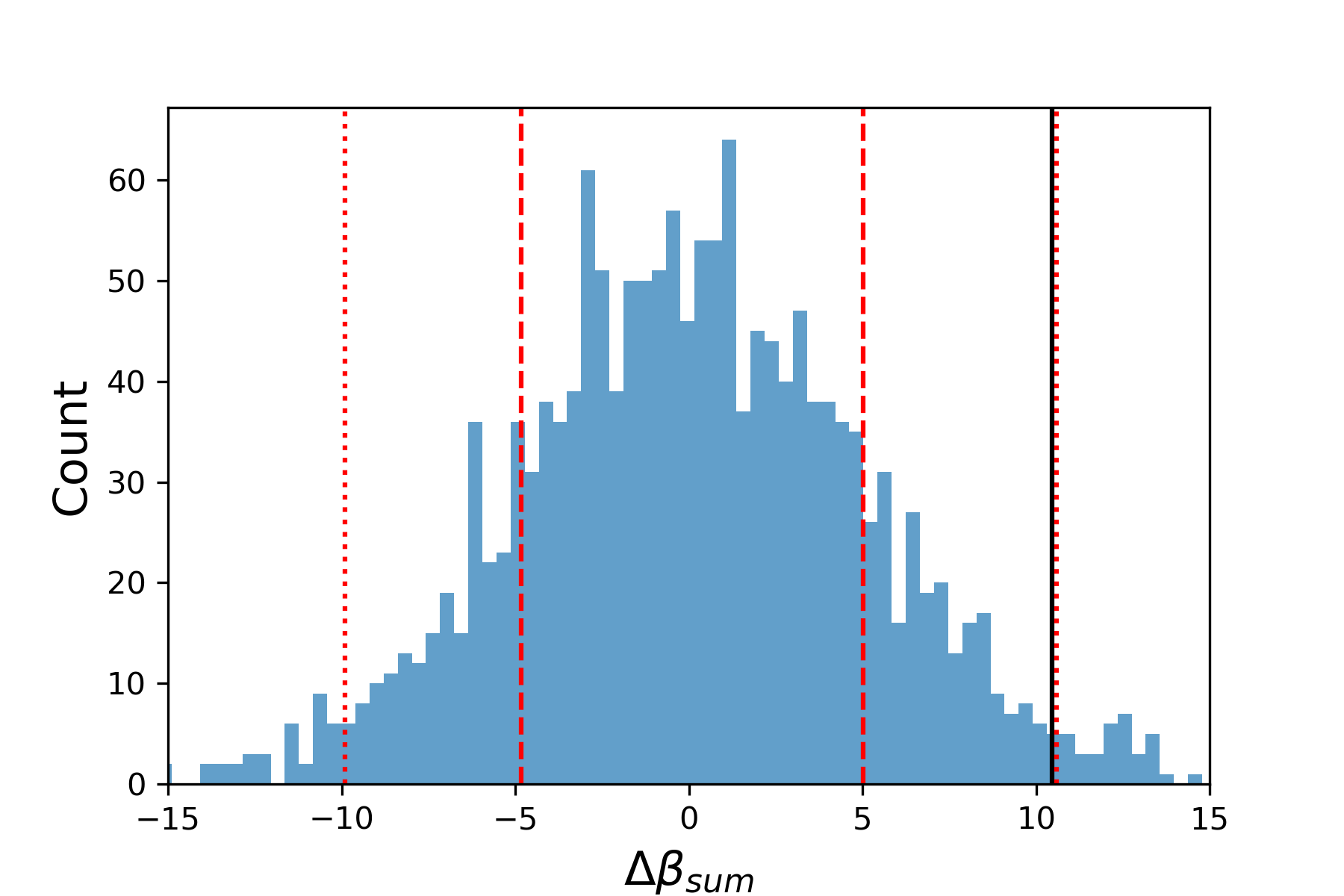}
\caption{The distribution of the weighted sum $\Delta \beta_{\textrm{sum}}$ for 1600 mock realisations of the real data from FIG. \ref{qso_dipole_z17}. The red dashed and dotted lines represent $1 \sigma$ and $2 \sigma $, respectively, while the value for the real data, $\Delta \beta_{\textrm{sum}} = 10.45$, corresponds to the black line. }
\label{sim_qso_subsample}
\end{figure}

We now return to the original data and FIG. \ref{qso_dipole}. For the full QSO sample in the range $0.7 < z \leq 7.5413$, we find $\Delta\beta_{\textrm{sum}} = 7.98$. In FIG. \ref{sim_qso}, we show the result of 1500 mock realisations of the data.  From 1500 mock realisations, we find a larger value of the weighted sum in 11 cases, which represents a probability of $p = 0.007$ or $2.4 \sigma$ for a single-sided Gaussian. This number can be compared with a $2.7 \sigma$ discrepancy if we only focused on the CMB dipole direction (see Table \ref{qso}.) The significance is evident from the position of the black line relative to the $1 \sigma$ and $2 \sigma$ confidence intervals in FIG. \ref{sim_qso}. For the more conservative subsample in FIG. \ref{qso_dipole_z17}, the corresponding value of the weighted sum is $\Delta\beta_{\textrm{sum}} = 10.45$. From 1600 mock realisations of flat $\Lambda$CDM data, we find larger values of $\Delta\beta_{\textrm{sum}}$ in 40 realisations. This represents a probability of $p = 0.025$, or $2 \sigma$, to get a larger emergent dipole in the CMB dipole direction simply as a statistical fluctuation. This number can be compared with the $2 \sigma$ discrepancy in the CMB dipole direction (see Table \ref{qso_z17}). The realisations are plotted in FIG. \ref{sim_qso_subsample}, and once again, the statistical significance is visually evident. 

\subsection{Caveats}
\label{sec:qso_caveats}
One caveat or feature of our QSO analysis is that it rests exclusively on flat $\Lambda$CDM. Since the QSO data returns large values of $\Omega_m$, and so too do HST SN \cite{Horstmann:2021jjg, Dainotti:2021pqg} as we explain in appendix \ref{sec:QSOHD}, it is expected that changes in the dark energy model, which primarily affect low redshifts,  do not affect $\beta, \gamma$ and this can be confirmed from Table 3 of Ref. \cite{Khadka:2020tlm}. In particular, observe that the $\beta$ errors do not increase as the model changes, which implies that the significance of the deviation we see in flat $\Lambda$CDM will not change across dark energy models. However, introducing curvature $\Omega_{k}$ causes $\beta$ to jump as is clear from a comparison of the ``Flat $\Lambda$CDM" and ``Non-flat $\Lambda$CDM" entries in Table 3 \cite{Khadka:2020tlm}. But this can be easily explained. 
As touched upon earlier, QSOs prefer smaller $D_{L}(z)$ values than Planck-$\Lambda$CDM. However, when one introduces curvature $\Omega_{k}$, for $\Omega_{k} < 0$, a sine function appears in the definition of $D_{L}(z)$ - see e.g. eq.(9) of \cite{Khadka:2020tlm} - that is bounded above by unity. Thus, QSO data can exploit this bound and saturate to lower $D_{L}(z)$ values. This can lead to turning points in $H(z)$, as is clear from some of the results in \cite{Khadka:2020tlm}.\footnote{See \cite{Colgain:2021beg} for comments on turning points in $H(z)$ and implications for the Null Energy Condition.} The jump in $\beta$ can nonetheless be simply explained by the additional freedom beyond $\Omega_{m}$, which the data gains to reduce $D_{L}(z)$, and increase $H(z)$, at higher $z$. 

\begin{figure}[htb]
\centering
  \includegraphics[width=80mm]{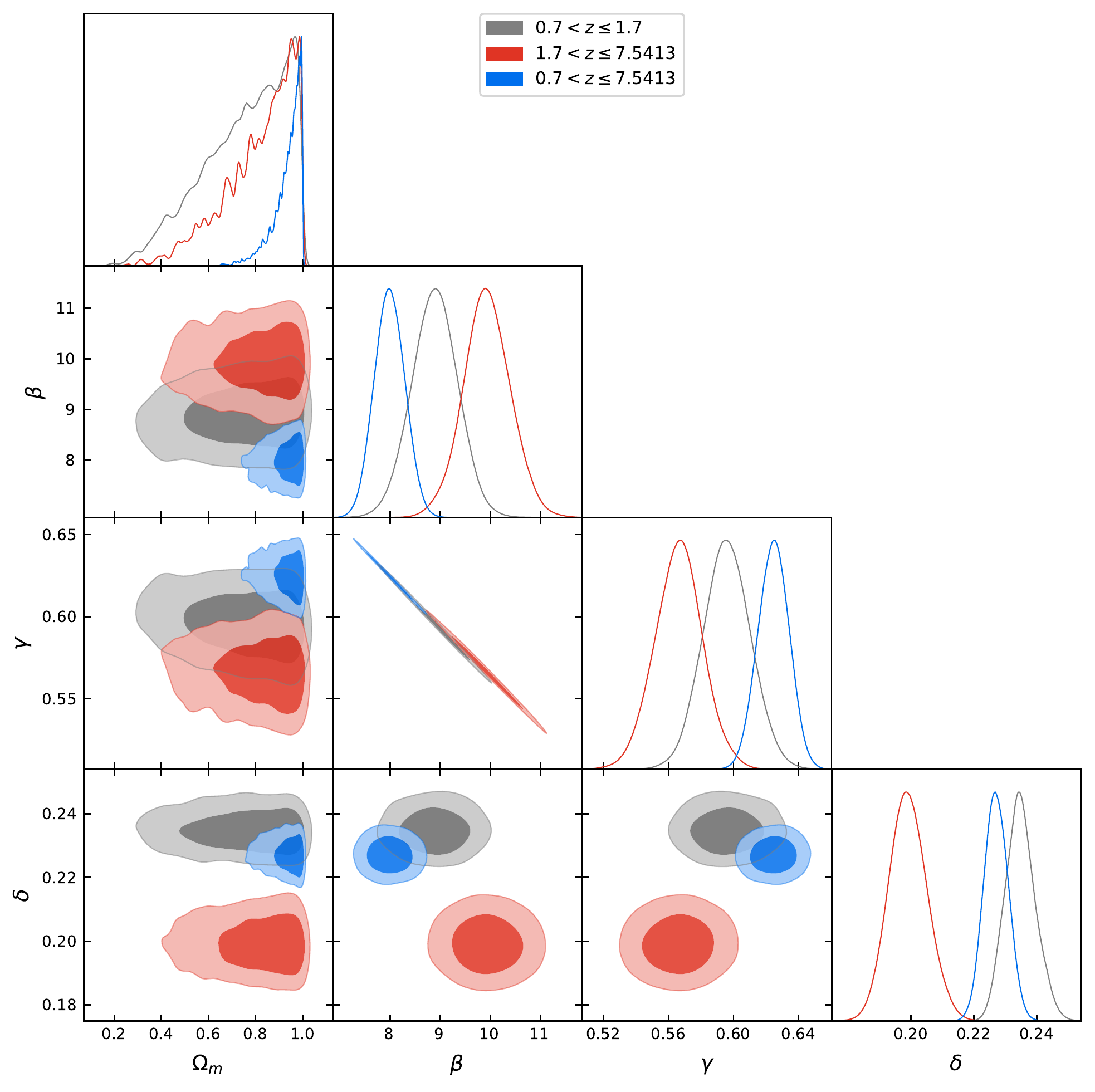}
\caption{Differences in the parameters $\Omega_{m}, \beta, \gamma$ and $\delta$ between low and high redshift samples. Here we made use of \textit{getdist} \cite{Lewis:2019xzd}.}
\label{lowz_highz}
\end{figure}

A secondary caveat concerns the observation within flat $\Lambda$CDM that low and high redshift QSOs lead to inconsistent values when combined. Using the redshift $z=1.7$ \footnote{This redshift choice is motivated by the findings of Ref. \cite{Khadka:2020tlm}, but is slightly higher than the $z=1.479$ value considered there.} as the border between low and high $z$, we confirm this in FIG. \ref{lowz_highz}. Evidently, the red (high $z$) region may be marginally discrepant with grey (low $z$), but our cut-off $z=1.7$ is low enough that the grey and blue regions (all $z$) are more or less consistent. Interestingly, the intrinsic dispersion $\delta$ drops considerably at higher redshifts. This lower $\delta$ at higher redshift is also evident in Table 3 of \cite{Khadka:2020tlm}. This is expected as sparser high redshift data possesses less scatter, and with less data, results become less robust. Ultimately, there may be some evolution of $(\beta, \gamma)$ with redshift and neither are strict constants. In appendix \ref{sec:QSOHD} we focus on redshift ranges below $z=1$, which we largely omit anyway, but it is clear that $(\beta, \gamma)$ evolution is pronounced at low redshifts, while becoming more gradual with increasing redshifts. Nevertheless, as is evident from FIG. \ref{lowz_highz} over extended redshift ranges some evolution may persist and correcting for this evolution will be important going forward \cite{Dainotti:2022rfz}. 

Here, it is useful to recall that relative to their status in the 1990s, SN, which we stress are only observed over a limited redshift range, have since been corrected for i) colour ii) shape and iii) host galaxy mass, so if QSOs follow suit, further corrections or sample selections will be warranted. That being said, as we show in appendix \ref{sec:QSOHD}, even uncalibrated QSOs identify a $\Lambda$CDM Universe at lower redshifts in line with SN. In other words, QSOs agree on some basic facts. However, the pertinent question is whether redshift evolution, namely a potential increasing $\beta$ with $z$ could mimic an orientational feature? This is possible if, when one splits the sample in hemispheres, the CMB dipole hemisphere, where $\beta$ is larger, shows QSO counts that increase from low to high redshifts. For maximum effect, the QSO counts in the opposite hemisphere would need to show the opposite trend, namely QSO counts that decrease with redshift. Instead, we find that QSO counts away from the CMB dipole are uniform in redshift in the range $0.7 < z < 1.7$, whereas QSO counts in the CMB dipole direction show a decreasing trend (see FIG. \ref{QSOcounts_redshift}), thereby hindering any tendency for redshift evolution to mimic a directional dependence.
%Whether any evolution in $(\beta, \gamma)$ is a physical feature or simply an artefact of high redshift data becoming sparse can be teased out with future QSO samples.  That being said, \red{
%there is no glaring inconsistency between grey and blue regions, so we are optimistic that both FIG. \ref{qso_dipole} and FIG. \ref{qso_dipole_z17} are tracking $H_0$ appropriately. \red{Naturally, it is good to be conservative and for this reason we quote results with the full QSO sample and a lower redshift subsample.}

In summary, we are seeing evidence for a higher value of $H_0$ within the flat $\Lambda$CDM model in the direction of the CMB dipole or aligned directions.  Since we treat both hemispheres equally, there is no obvious bias. Indeed, FIG. \ref{qso_dipole} and FIG. \ref{qso_dipole_z17} are reminiscent of similar features in strong lensing time delay \cite{Krishnan:2021dyb} and Type Ia SN \cite{Krishnan:2021jmh}. The naive interpretation is that there is a dipole anisotropy in the matter density of the Universe as traced by QSOs. This is consistent with the observation that there is a mismatch in the magnitude of the cosmic dipole between QSOs and CMB \cite{Secrest:2020has}, but here (partial) sky coverage is not a concern.  

\section{GRBs}
\label{sec:GRB}

\subsection{Uncalibrated GRBs}
\begin{figure}[htb]
\centering
  \includegraphics[width=80mm]{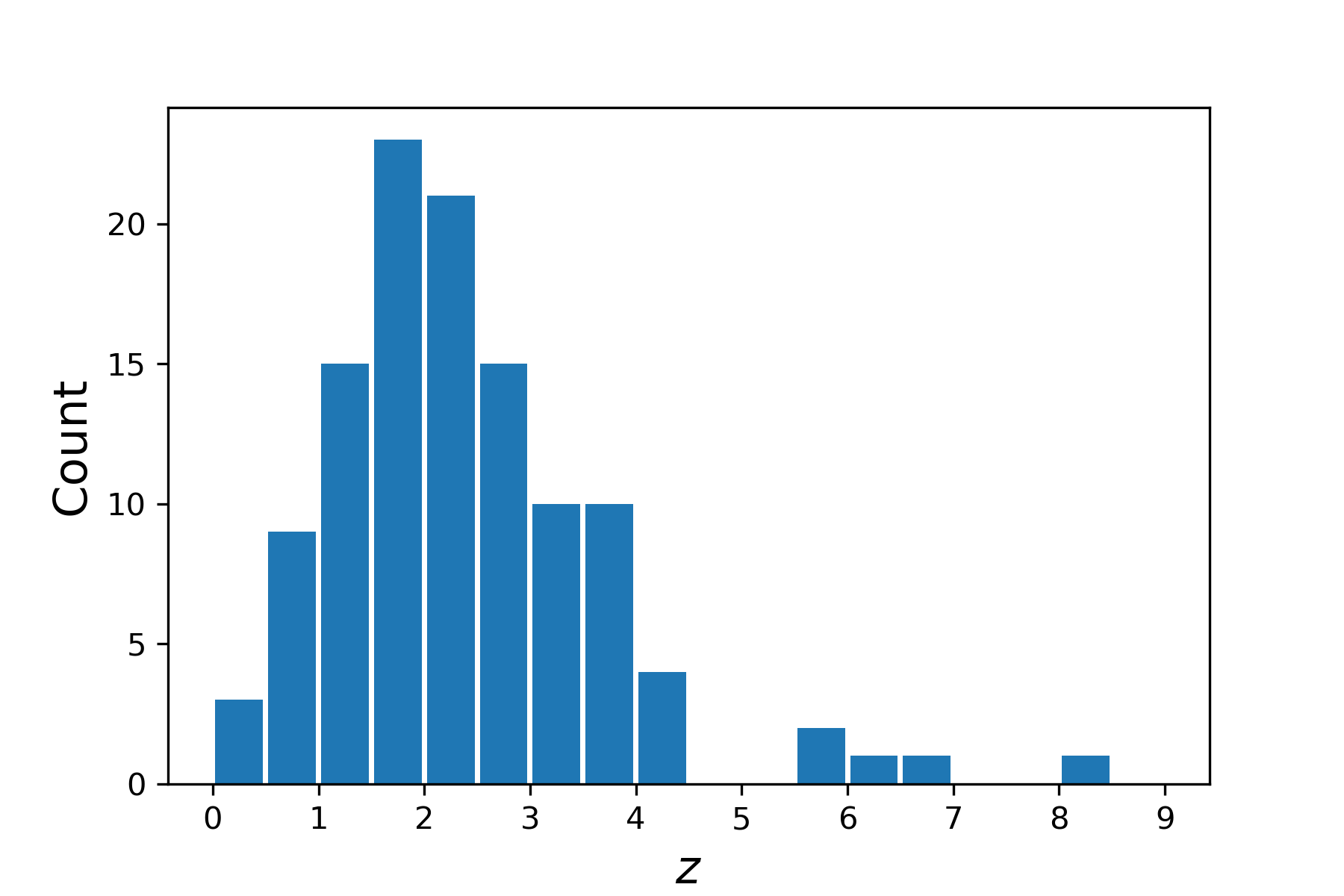}
\caption{Distribution of 118 GRBs with redshift $z$ in intervals of $\Delta z = 0.5$.}
\label{grb_count}
\end{figure}

\begin{figure}[htb]
\centering
  \includegraphics[width=70mm]{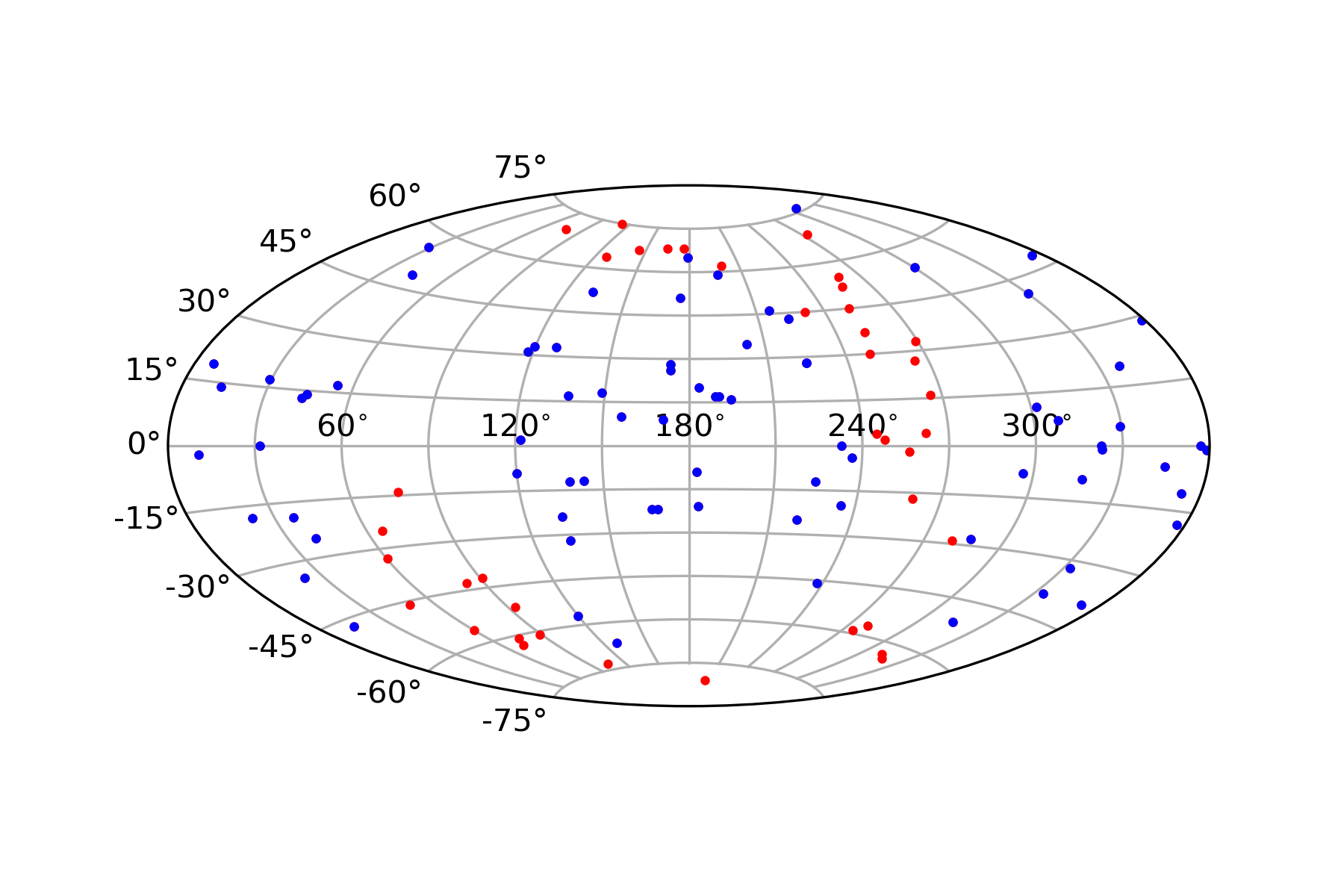}
\caption{Distribution of the 118 GRBs \cite{Khadka:2021dsz} on the sky.}
\label{grb_loc}
\end{figure}

Our complementary analysis here is guided by the findings  in \cite{Khadka:2021dsz}, where assuming the Amati correlation \cite{Amati:2008hq}, a compilation of 118 GRBs in the redshift range $ 0.3399 \leq z \leq 8.2$ with small enough intrinsic dispersion was identified. See FIG. \ref{grb_count} for the sample's redshift distributon and FIG. \ref{grb_loc} for the sample's orientation on the sky. The relatively low scatter suggests that this sample may currently be the optimal one for cosmological studies. Given the improvement relative to previous samples, e. g. \cite{Demianski:2016zxi}, it is interesting to see if GRBs, which are also high redshift probes, exhibit the same feature as QSOs. Throughout, it is worth bearing in mind that GRB samples are smaller, so statistically less powerful, while also possessing greater intrinsic scatter. We include them in our analysis on the grounds that like QSOs, they are considered promising cosmological probes at higher redshifts \cite{Schaefer:2007s, Wang:2007, Amati:2008hq, Capozziello:2008, Dainotti:2008, Izzo:2009, Amati:2013, Wei:2014, Izzo:2015, Tang:2019}.

Recall that the Amati correlation relates the spectral peak energy
in the GRB cosmological rest-frame $E_{\textrm{p,i}}$  and the isotropic energy $E_{\textrm{iso}}$: 
\be
\label{eq1}
\log_{10} E_{\textrm{iso}} = \alpha + \beta \log_{10} E_{\textrm{p,i}},  
\ee
where $\alpha$ and $\beta$ are free parameters. Neither $E_{\textrm{iso}}$ nor $E_{\textrm{p,i}}$ are observed quantities. The latter is related to the similar quantity in observer's frame as %$E_{\textrm{p}}^{\textrm{obs}}$ through a simple redshift-dependent factor, 
$E_{\textrm{p,i}} = E_{\textrm{p}}^{\textrm{obs}} (1+z)$ and the former depends on the cosmology through the luminosity distance $D_{L}(z)$ and the measured bolometric fluence $S_{\textrm{bolo}}$ \cite{Demianski:2016zxi},
\be
E_{\textrm{iso}} = 4 \pi D_{L}^2 (z) S_{\textrm{bolo}} (1+z)^{-1}.
\ee
%yielding
%\be\label{eq-GRB}
%\log_{10} S_{\textrm{bolo}} = \alpha + \beta \log_{10} E_{\textrm{p}}^{\textrm{obs}}+(\beta+1)\log_{10}(1+z)-\beta \log_{10} (4\pi D^2_L). \nonumber 
%\ee
Once again we focus on flat $\Lambda$CDM and to address scatter in the GRB data, an intrinsic dispersion parameter $\delta$ is introduced. In line with previous analysis, we fix $H_0 = 70$ km/s/Mpc and adopt $\Omega_m, \alpha, \beta$ and $\delta$ as the free parameters. The best-fit values are identified by extremising the following likelihood function: 
\be
\label{L2}
\mathcal{L} = - \frac{1}{2} \sum_{i=1}^{N} \left[ \frac{ \left(\log_{10} E_{\textrm{iso}, i} - (\alpha + \beta \log_{10} E_{\textrm{p}, i} )\right)^2}{s^2_{i}} + \ln (2 \pi s_i^2) \right], 
\ee 
where $N$ is the number of GRBs. In addition, $s_i$ depends on the $S_{\textrm{bolo}}, E_{\textrm{p,i}}$, the corresponding errors $\sigma_{S_{\textrm{bolo}}}, \sigma_{E_{\textrm{p,i}}}$ and the intrinsic dispersion, 
\be
s_i^2 = \left(  \frac{\sigma_{S_{\textrm{bolo}, i}}}{S_{\textrm{bolo}, i} \ln (10)} \right)^2 +  \beta^2 \left( \frac{\sigma_{E_{\textrm{p}, i}}}{E_{\textrm{p}, i} \ln (10)} \right)^2 + \delta^2. 
\ee

We see from equation (\ref{eq1}) that the degeneracy between $H_0$ and $\alpha$ means that an \textit{increase} in $H_0$ corresponds to an \textit{increase} in $\alpha$. Therefore, relative to QSOs, the degeneracy is the same. 

\begin{figure}[htb]
\centering
  \includegraphics[width=80mm]{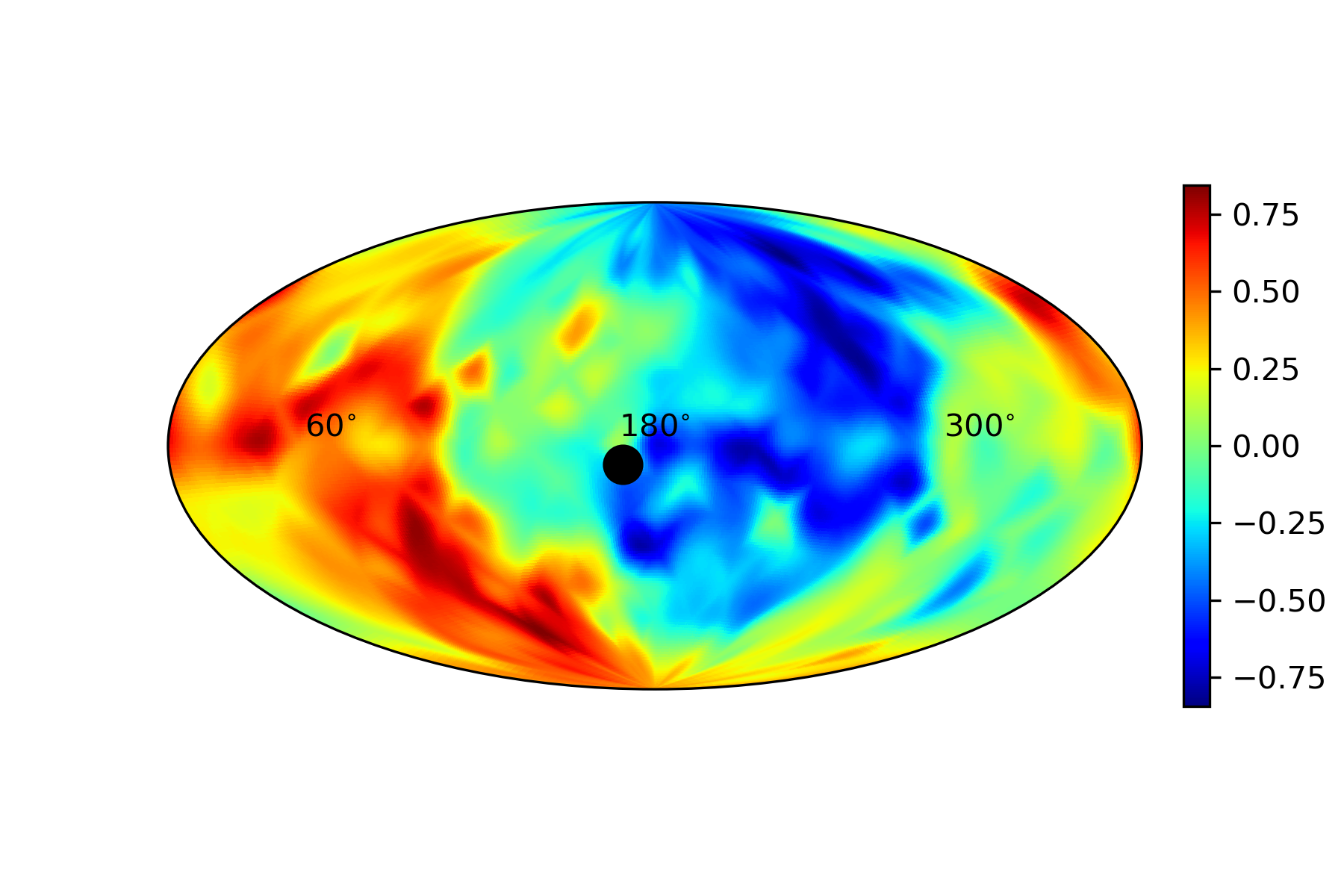}
\caption{Variations of the best-fit $\alpha$ parameter in respective hemispheres as (RA, DEC) values for the GRBs are scanned over. The black dot denotes the CMB dipole. %Note that a higher (lower) value $\alpha$ corresponds to lower (higher) value $H_0$.
}
\label{grb_dipole}
\end{figure}

The methodology is the same as section \ref{sec:QSO} and the result of the scan over (RA, DEC) can be found in FIG. \ref{grb_dipole} and Table \ref{grb}, where we record best-fit parameters for the CMB dipole direction and the direction of greatest $\Delta \alpha$. We suppress $\delta$ in Table \ref{grb} as it consistently returns values in the vicinity of $\delta \approx 0.4$. Given that we have an order of magnitude fewer GRBs than QSOs, it is not surprising to see that any deviation in $\Delta \alpha$ is not so pronounced. From Table \ref{grb}, we find that the discrepancy in $\alpha$ across hemispheres is $0.5 \sigma$ lower for the CMB dipole and $1.6 \sigma$ for the direction of  maximum $\Delta \alpha$, but this maximum in $\Delta \alpha$, corresponding to a maximum in $\Delta H_0/H_0$ is actually in the hemisphere away from the CMB dipole. Thus, we arrive at the opposite result to the QSOs, namely with the sample of 118 GRBs \cite{Khadka:2021dsz}, which have been compiled and presented as the optimal current sample for cosmological studies, we find that $H_0$ is in fact smaller in the CMB dipole direction.

\begin{table}[htb]
\centering 
\begin{tabular}{c|c|c|c|c}
\rule{0pt}{3ex} (RA, DEC)& Hemisphere & $ \Omega_{m}$ & $ \alpha$ & $\beta$ \\
\hline 
\rule{0pt}{3ex} \multirow{ 4}{*}{CMB dipole} &  \multirow{ 2}{*}{N} & $1$ & $49.83$ & $1.17$  \\
\rule{0pt}{3ex} & &  $0.67^{+0.23}_{-0.30}$ & $49.91^{+0.40}_{-0.36}$ & $1.20^{+0.13}_{-0.15}$ \\
\cline{3-5}
\rule{0pt}{3ex}  & \multirow{ 2}{*}{S} &  $0.38$ & $50.22$ & $1.08$ \\
\rule{0pt}{3ex} & &  $0.52^{+0.32}_{-0.27}$ & $50.17^{+0.30}_{-0.28}$ & $1.08^{+0.10}_{-0.10}$ \\
\hline 
%\rule{0pt}{3ex} \multirow{ 4}{*}{\red{$(96^{\circ}, -64.3^{\circ})$}} 
%\rule{0pt}{3ex}  & \multirow{2}{*}{N} & \red{$0.61$} & %\red{$50.48$} & \red{$0.98$} \\ 
%\rule{0pt}{3ex}  & & $0.64^{+0.25}_{-0.30}$ & $49.72^{+0.36}_{-0.34}$ & $1.24^{+0.12}_{-0.12}$  \\
%\cline{3-5}
%& \multirow{ 2}{*}{S} & \red{$1$} & \red{$49.64$} & \red{$1.21$} \\
%\rule{0pt}{3ex} & & $0.60^{+0.27}_{-0.29}$ & $50.49^{+0.34}_{-0.34}$ & $0.98^{+0.13}_{-0.12}$ \\
\rule{0pt}{3ex} \multirow{ 4}{*}{$(96^{\circ}, -64.3^{\circ})$} & \multirow{ 2}{*}{N} & $0.72$ & $50.46$ & $0.96$ \\
\rule{0pt}{3ex} & & $0.60^{+0.27}_{-0.29}$ & $50.49^{+0.34}_{-0.34}$ & $0.98^{+0.13}_{-0.12}$ \\
\cline{3-5}
\rule{0pt}{3ex}  & \multirow{2}{*}{S} & $1$ & $49.61$ & $1.23$ \\ 
\rule{0pt}{3ex}  & & $0.64^{+0.25}_{-0.30}$ & $49.72^{+0.36}_{-0.34}$ & $1.24^{+0.12}_{-0.12}$  \\
\end{tabular}
\caption{Best-fit values of the parameters from both extremisation and MCMC analysis of the likelihood (\ref{L2}) for GRBs in the redshift range $0.3399 \leq z \leq 8.2$. Note, we quote the maximum value of $\Delta \alpha$ corresponding to the maximum value of $\Delta H_0/H_0$.}
\label{grb}
\end{table}

Given the disagreement between QSOs and GRBs, 
%it is admittedly a little surprising that FIG. \ref{grb_dipole} shows good \textit{qualitative} agreement with FIG. \ref{qso_dipole}. However, this agreement may be deceiving and 
it is best once again to turn to simulations in order to ascertain the significance of the uncalibrated GRB result. Our methodology is as in section \ref{sec:QSO}, we  use the  weighted sum (\ref{weighted_sum}) except that here we track deviations in $\Delta \alpha$, $\Delta\alpha_{\textrm{sum}} = \sum_{i=1}^{25} w_i\ \Delta \alpha_i.$ 
Once again, we perform an MCMC exploration of the original sample to identify representative values of $(\Omega_{m}, \alpha, \beta, \delta$), and for each entry in the MCMC chain, we mock up new values of $E_{\textrm{p,i}}$ by utilising the error $\sigma_{E_{\textrm{p,i}}}$, before generating values of $S_{\textrm{bolo}}$, so that $E_{\textrm{iso,i}}$ conform to the relation (\ref{eq1}) with standard deviation $s_i$. For a number of these mock values, we have confirmed that the mock gives reasonable results in the sense that one is typically within the $1 \sigma $ confidence intervals of the best-fit parameters of $(\Omega_{m}, \alpha, \beta, \delta)$, as defined by the overall MCMC chain. Throughout, we have discarded any mocks with unphysical values of $E_{\textrm{p,i}}$, namely $E_{\textrm{p, i}} < 0$. This did not pose a problem with the earlier QSO mocks, but was a more prevalent problem here. 

For the weighted sum evaluated on the real configuration in FIG. \ref{grb_dipole}, we find $\Delta\alpha_{\textrm{sum}} = -0.92$. As expected, we see that the weighted sum returns a negative value consistent with lower values of both $\alpha$ and $H_0$ in the CMB dipole hemisphere. Nevertheless, from 1432 mock realisations of the data, we find $1 \sigma$ confidence intervals of $-2.09 < \Delta\alpha_{\textrm{sum}} < 2.44$ for the weighted sum.  %In particular, we find a lower value of $\Delta\alpha _{\textrm{sum}}$ in 433 cases, which means that the probability of a similar or lower value of $\Delta\alpha _{\textrm{sum}}$ arising simply by chance is $p = 0.30$. 
Therefore, the configuration in FIG. \ref{grb_dipole} is consistent with a random statistical fluctuation within the flat $\Lambda$CDM model. In other words, we have some conflict between QSO and GRB samples, but FIG. \ref{grb_dipole} is not significant within flat $\Lambda$CDM and is fully consistent with an isotropic Universe. 

While the relatively small sample size is a factor, it is also possible that scatter is playing a role.
It should be noted that the intrinsic dispersion in the GRB sample, $\delta \sim 0.4$, is considerably larger than the QSO sample, $\delta \sim 0.24$. While one could attempt to reduce the scatter by hand by analysing the residuals from best-fit cosmologies and removing outliers, %one can instead turn to orientation. 
there is another notable difference between the QSO and GRB samples. As is clear from FIG. \ref{qso_loc} and FIG. \ref{grb_loc}, the GRBs are distributed on the sky in a more isotropic fashion. For this reason, it is interesting to remove GRBs so that the resulting subsample bears a closer resemblance to the clearly anisotropic QSO distribution. It should be stressed that since we are working within the FLRW paradigm, there is no penalty from removing observables based on orientation. 

As a result, it is a valid exercise to consider the weight \eqref{weight} and steadily remove GRBs with lower values of the absolute magnitude of $w_i$. This removes GRBs that are less closely aligned or misaligned with the CMB dipole. Conducting this exercise for the original sample in steps of $|\Delta w_i| = 0.1$ through to only GRBs with $|w_i| > 0.5$, we did not find any pronounced change in the dipole. Moreover, by calculating $\Delta\alpha_{\textrm{sum}}$ we confirmed through simulations that any configuration was consistent with a statistical fluctuation within the flat $\Lambda$CDM model. This exercise serves as a prelude to similar analysis with calibrated GRBs in the next section, where we find different results.

\subsection{Calibrated GRBs}
\label{sec:calibratedGRB}
In this section we analyse a second compilation of 162 GRBs in the redshift range $0.03351 \leq z \leq 9.3$ that have been calibrated by Type Ia SN \cite{Demianski:2016zxi}. We remove all GRBs below $z = 0.3$ and this leaves us with 158 GRBs, which should be deep enough in redshift to share the same frame as the CMB. Since the data is in distance moduli, there are no nuisance parameters e. g. $\alpha, \beta, \delta$, and we can directly fit $(H_0, \Omega_{m})$ to the distance moduli.

\begin{figure}[htb]
\centering
  \includegraphics[width=80mm]{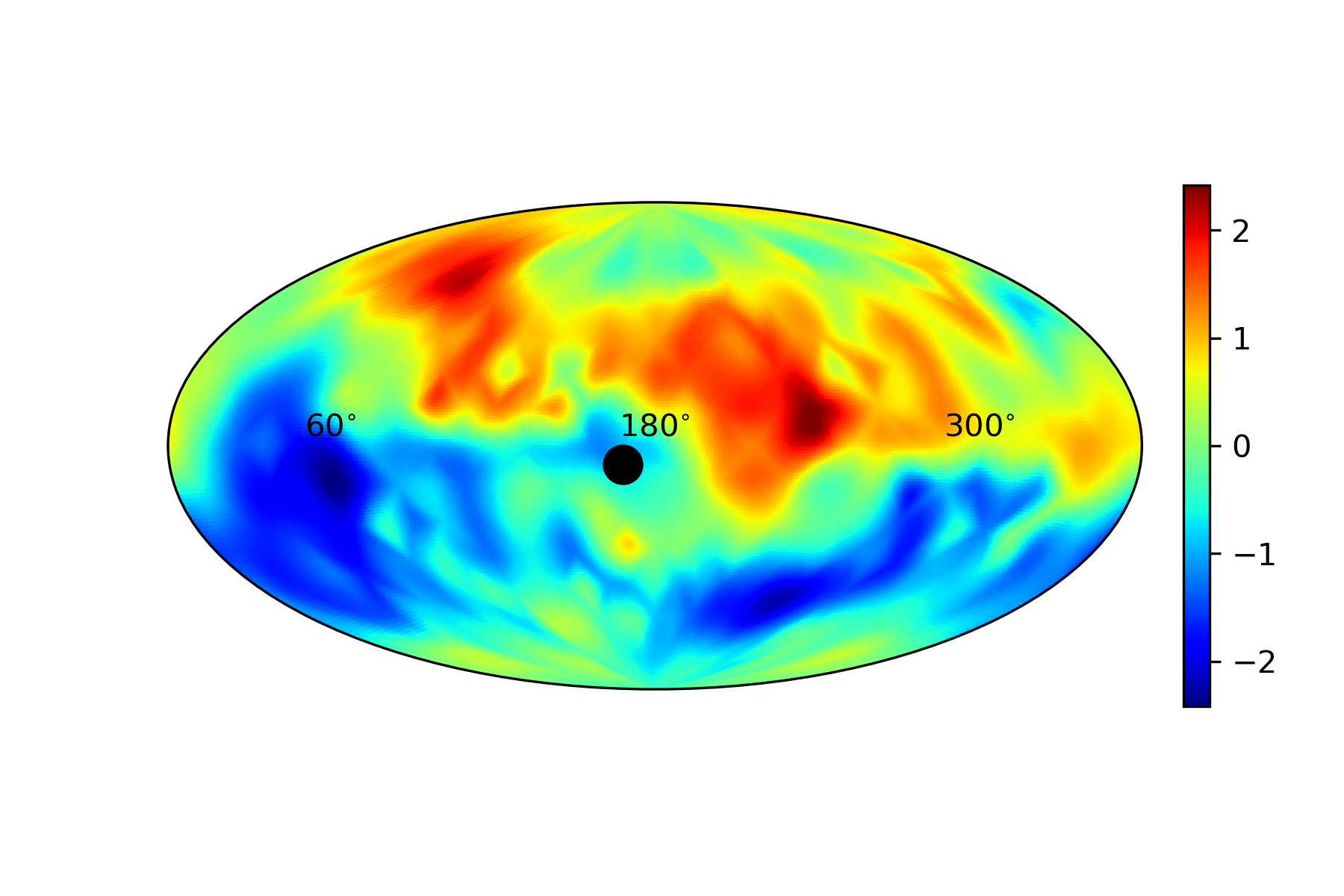}
\caption{Variations of (\ref{sigma}) across the sky for 158 GRBs of
\cite{Demianski:2016zxi} in the redshift range $0.3 < z \leq 9.3$. The black dot denotes the CMB dipole.}
\label{demianski1}
\end{figure}

Repeating the process outlined in the text with the full data set, while focusing on the (unweighted) hemisphere split quantity \cite{Krishnan:2021jmh}, 
\be
\label{sigma}
\sigma := (H_0^{N} - H_0^{S})/\sqrt{(\delta H_0^N)^2+(\delta H_0^S)^2}, 
\ee 
one arrives at FIG. \ref{demianski1}. Note that in contrast to earlier plots, here the colour map is recording the significance of the discrepancy and not just the discrepancy. Obviously, this plot shows little or no correlation and is inconclusive. Moreover, in the CMB dipole direction $H_0$ is actually lower, thus contradicting our hypothesis. This is consistent with the uncalibrated GRB sample. Nevertheless, we are encountering $\sim 2 \sigma$ displacements at certain points on the sky. 

\begin{figure}[htb]
\centering
  \includegraphics[width=80mm]{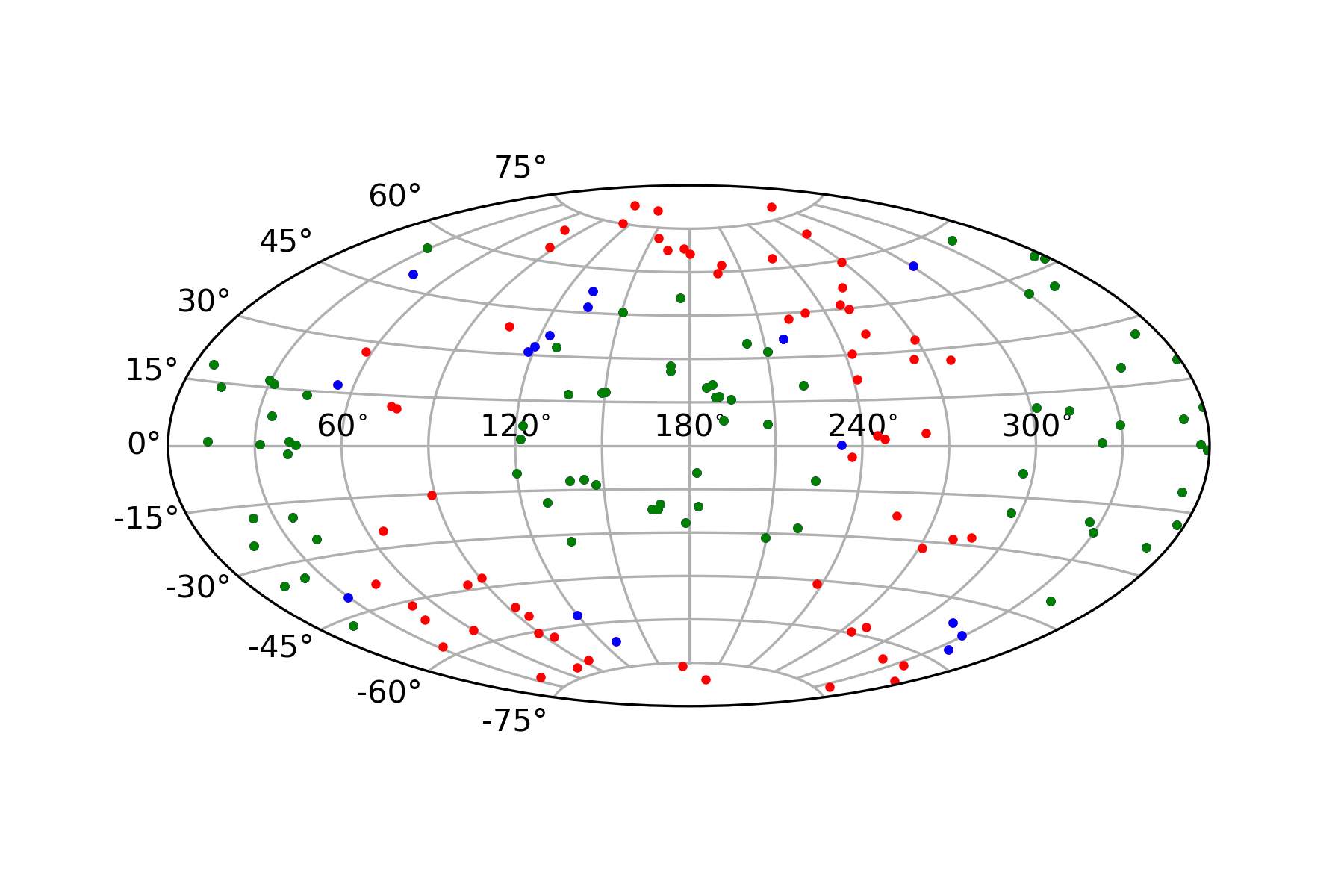}
\caption{Location of 158 GRBs in the redshift range $0.3 < z \leq 9.3$ \cite{Demianski:2016zxi} on the sky. The green dots denote a subsample of 78 GRBs with closest alignment/misalignment with the CMB dipole. Including the blue dots and red dots, one has 95 GRBs and 158 GRBs, respectively.}
\label{cal_grb_loc}
\end{figure}

In FIG. \ref{cal_grb_loc} we record the location of the GRBs on the sky. In contrast to the QSOs in FIG. \ref{qso_loc} and smaller flux GRB sample in FIG. \ref{grb_loc},  one can see that the calibrated GRBs offer better sky coverage. 
%It is also worth stressing again that GRBs have a larger intrinsic dispersion, $\delta \sim 0.4$, relative to QSOs, $\delta \sim 0.23$. The other noticeable difference is the bunching in the QSO data away from the galactic plane. 
While the GRBs show no conclusive evidence for a dipole from FIG. \ref{demianski1}, at least within an FLRW setup, as we have explained, one has the freedom to play with orientation so that  the sky orientation more closely corresponds to the QSO distribution in FIG. \ref{qso_loc}. Such changes are not expected to make a difference. However, it should be clear that the QSOs are more closely (mis)aligned with the CMB dipole direction, and if there is an anisotropy, as suggested by existing discrepancies in the cosmic dipole \cite{Blake:2002gx, Singal:2011dy, Gibelyou:2012ri, Rubart:2013tx, Tiwari:2015tba, Colin:2017juj, Bengaly:2017slg, Siewert:2020krp, Secrest:2020has, Singal:2021crs, Singal:2021kuu}, then we can ``zoom" in on this direction by removing poorly (mis)aligned GRBs. Thus, as we did in the last section, we can remove GRBs with smaller absolute values of the weight (\ref{weight}). 
%and focus on GRBs that are more closely aligned or misaligned with the CMB dipole direction. 
Given the weight (\ref{weight}), one can define a weighted sum 

\be\label{sigma-sum}
\sigma_{\textrm{sum}} = \sum_{i}^{25} w_i \sigma_i,
\ee
where $\sigma_i$ is defined in \eqref{sigma}, and repeat the compression of heat maps into a number through the weighted sum. 
%It appears that $\sigma_{\textrm{sum}}$ is a better measure for anisotropy than $\Delta\alpha_{\textrm{sum}}$.

\begin{figure}[htb]
\centering
  \includegraphics[width=80mm]{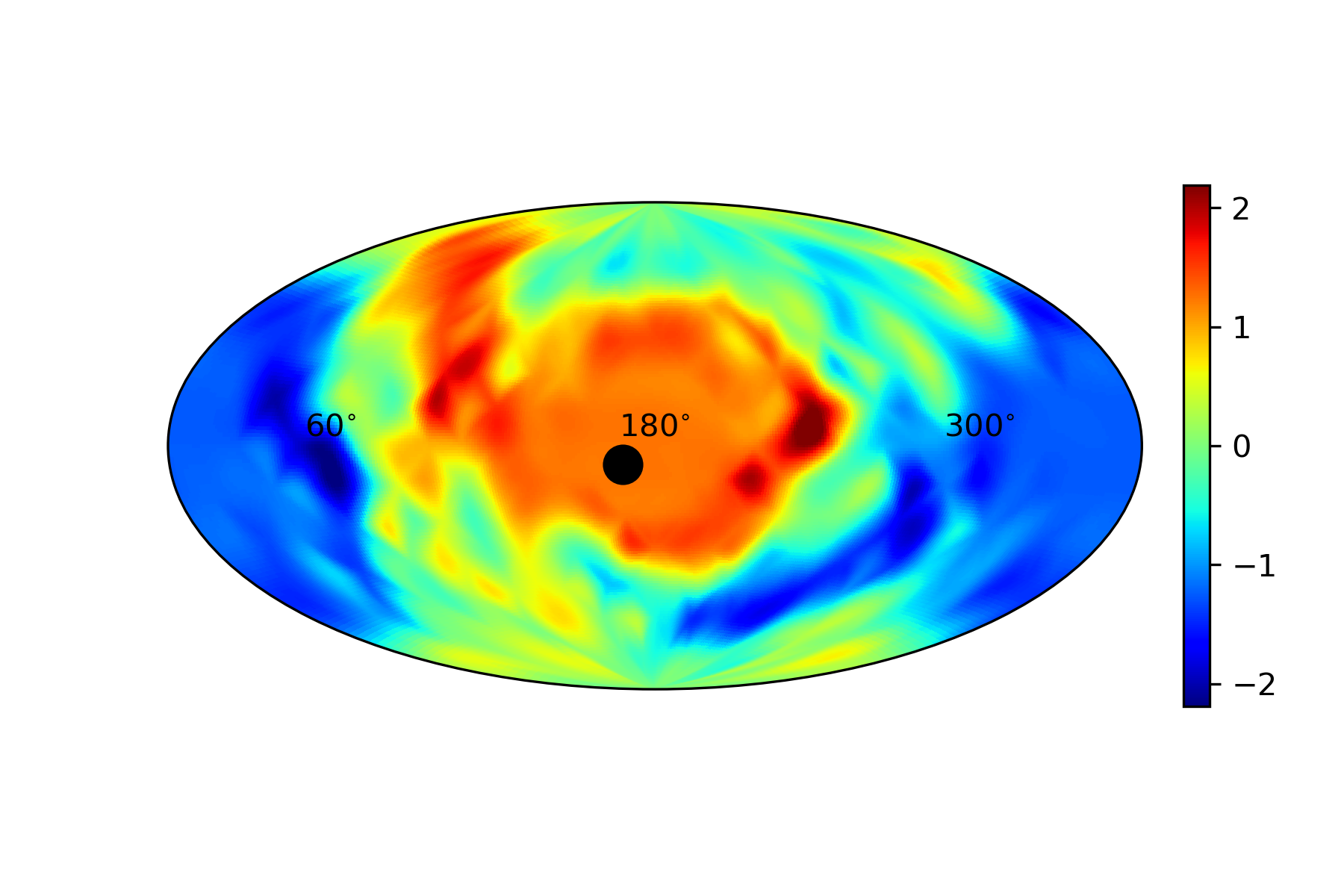}
\caption{Variations of (\ref{sigma}) across the sky for a 95 GRB subsample of the GRBs of Ref. \cite{Demianski:2016zxi}. We have removed GRBs less closely aligned or misaligned with the CMB dipole direction with $|w_i|<0.4$. The black dot denotes the CMB dipole.}
\label{demianski2}
\end{figure}

For the original 158 data points, we find that $\sigma_{\textrm{sum}} = - 0.64$, thus reinforcing that, if anything, $H_0$ is lower in the CMB dipole direction, thereby contradicting the working hypothesis. However, once the GRBs with $|w_i| < 0.4$ have been removed, this number changes dramatically to $\sigma_{\textrm{sum}} = 6.56$. The corresponding variation in $H_0$ with an interpolation based on the original $31 \times 15$ grid is shown in FIG. \ref{demianski2}. The plot is in line with expectations once one notices that when we remove smaller values of $|w_i|$, we are left with fewer GRBs away from the CMB dipole axis and this explains the expansion in the light green regions where the discrepancy in $H_0$ is negligible. It should be stressed that although we have trimmed the data set to get this pronounced dipole, some of the features are still evident in the original plot, FIG. \ref{demianski1}, including the dark red region at (RA, DEC) $\sim (240^{\circ}, 10^{\circ})$ and the extension of the red ``arm" towards the north pole.

\begin{figure}[htb]
\centering
  \includegraphics[width=80mm]{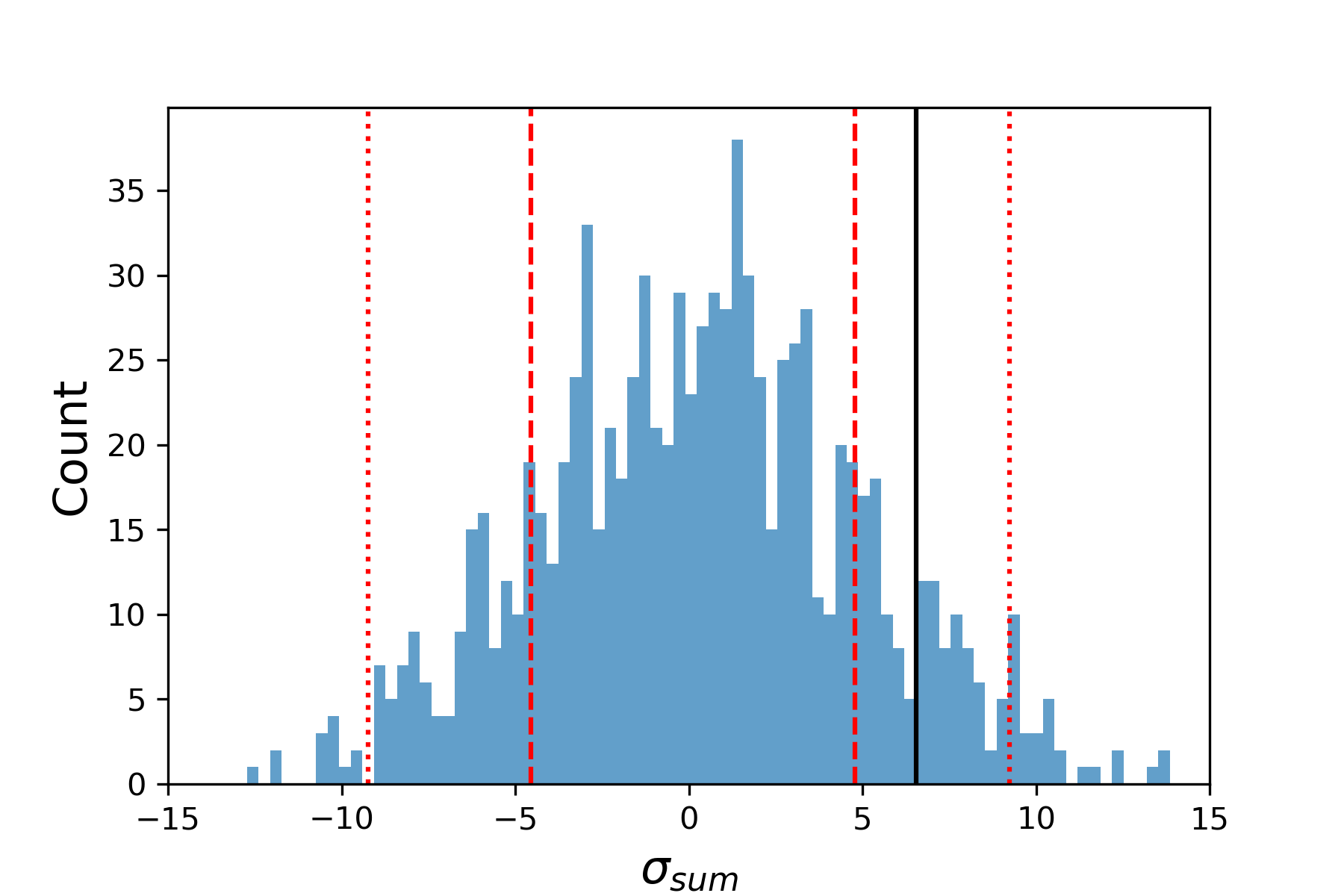}
\caption{The distribution of the weighted sum $\sigma_{\textrm{sum}}$ for 2616 mock realisations of the real data from FIG. \ref{demianski2}. The red dashed and dotted lines represent $1 \sigma$ and $2 \sigma $, respectively, while the value for the real data, $\sigma_{\textrm{sum}} = 6.56$ corresponds to the black line. }
\label{sim_cal_grb_w04}
\end{figure}

We can now establish the significance of FIG. \ref{demianski2} by mocking up data consistent with flat $\Lambda$CDM using the same 95 data points. Concretely, we adopt the redshifts and errors on the distance moduli, but mock up the distance moduli from a normal distribution in $(H_0, \Omega_{m})$ around the best-fit values, $H_0 = 73.01 \pm 11.81$ km/s/Mpc and $\Omega_{m} = 0.23 \pm 0.19$. From 2616 mock realisations, as illustrated in FIG. \ref{sim_cal_grb_w04}, we find 238 with larger values of $\sigma_{\textrm{sum}}$. Thus the probability of getting a more pronounced discrepancy in $H_0$ in the CMB dipole direction is $p = 0.091$. This represents $1.3 \sigma$ for a single-sided Gaussian, which is consistent with the solid black line in FIG. \ref{sim_cal_grb_w04}. 

\begin{figure}[htb]
\centering
  \includegraphics[width=80mm]{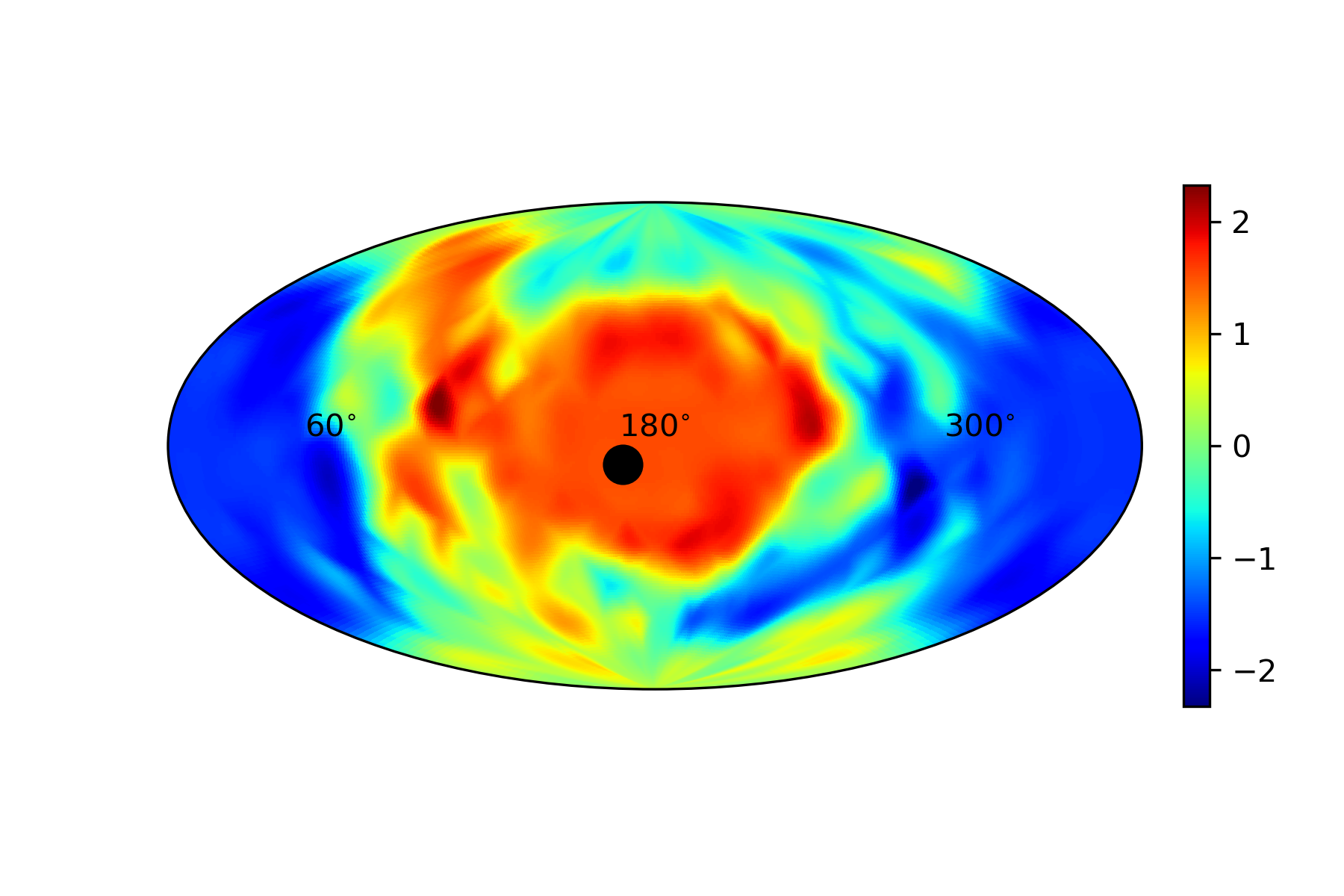}
\caption{Variations of (\ref{sigma}) across the sky for a 78 GRB subsample of the GRBs of Ref. \cite{Demianski:2016zxi}. We have removed GRBs less aligned or misaligned with the CMB dipole with $|w_i| < 0.5$. The black dot denotes the CMB dipole.}
\label{demianski3}
\end{figure}

\begin{figure}[htb]
\centering
  \includegraphics[width=80mm]{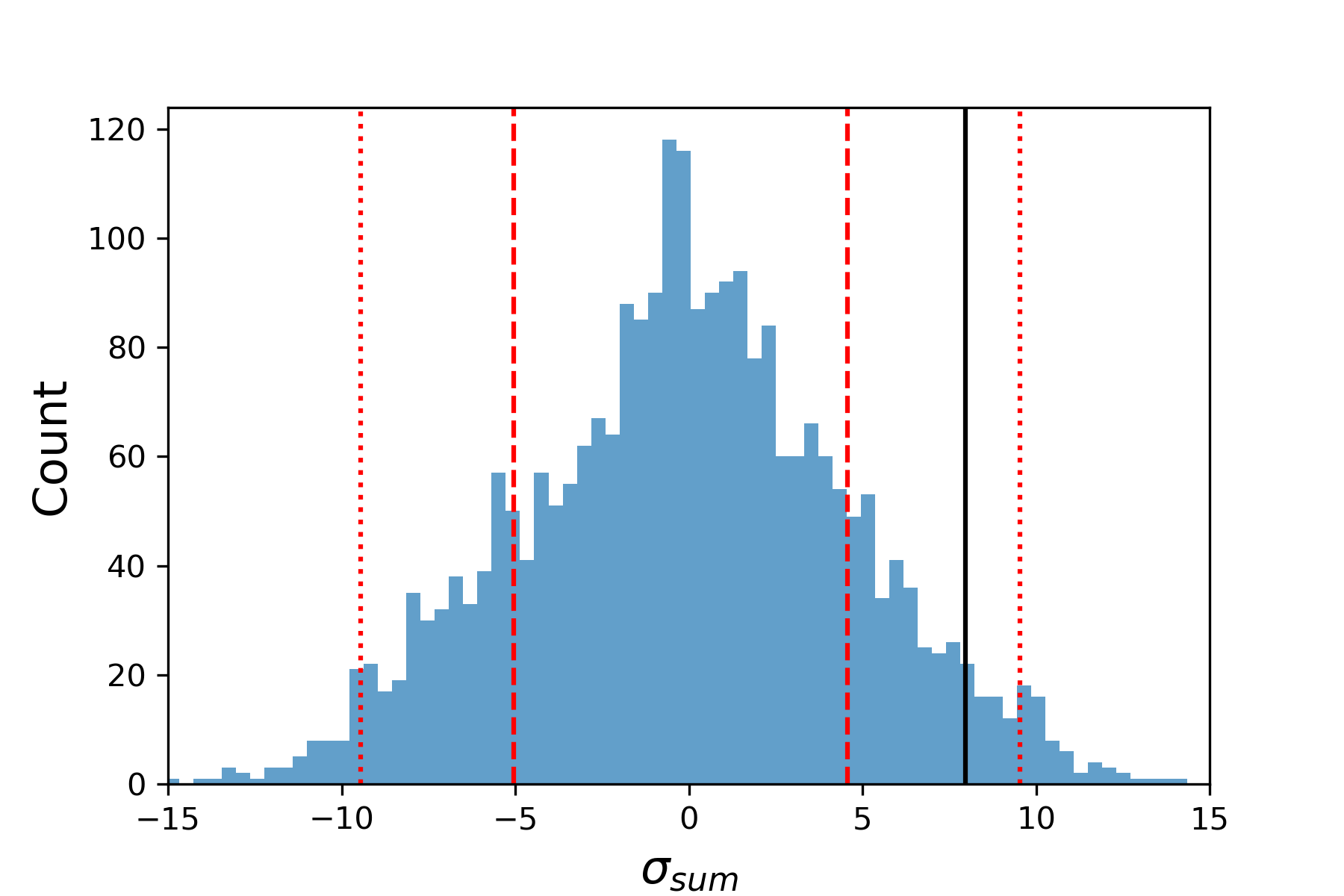}
\caption{The distribution of the weighted sum $\sigma_{\textrm{sum}}$ for 2575 mock realisations of the real data from FIG. \ref{demianski3}. The red dashed and dotted lines represent $1 \sigma$ and $2 \sigma $, respectively, while the value for the real data, $\sigma_{\textrm{sum}} = 7.98$ corresponds to the black line. }
\label{sim_cal_grb_w05}
\end{figure}

We can consider a further cut by removing GRBs with $|w_i| < 0.5$, which leaves us with 78 GRBs. From FIG. \ref{demianski3} we can see that in contrast to FIG. \ref{demianski2}, the red region has become a little deeper in colour. Using the reduced grid and weighted sum over 25 points in the same hemisphere as the CMB dipole direction, the sum increases from $\sigma_{\textrm{sum}} = 6.56$ to $\sigma_{\textrm{sum}} = 7.98$. Clearly, the emergent dipole has become more pronounced. Once again, we identify the best-fit parameters, $H_0 = 79.57 \pm 12.95$ km/s/Mpc and $\Omega_{m} = 0.23 \pm 0.16$ and construct 2575 mock realisations of the data. We find that 121 of these realisations lead to larger values of $\sigma_{\textrm{sum}}$. As a result, the probability of a similar weighted sum arising by chance from within flat $\Lambda$CDM is $p = 0.047$, or $1.7 \sigma$ for a single-sided Gaussian. The mocks are shown in FIG. \ref{sim_cal_grb_w05}.

In summary, we have seen how a relatively strong dipole can emerge from a calibrated GRB sample as the GRBs poorly (mis)aligned with the CMB dipole direction are removed. It should be stressed that while there may be hidden correlations, and this point certainly deserves further study, \textit{a priori} since we are working within an FLRW framework, the removal of GRBs in certain directions is not expected to greatly change the results. Nevertheless, as we have seen, once the poorly (mis)aligned GRBs are removed, we see that the sample comprises higher (lower) values of $H_0$ in directions aligned (misaligned) with the CMB dipole. The expanding light green regions in FIG. \ref{demianski2} and FIG. \ref{demianski3} point to the same fact, namely in directions away from the CMB dipole, where we have removed GRBs, any differences in $H_0$ are minimal. Evidently, this is because one may be picking GRBs pretty evenly from samples of GRBs that prefer higher and lower $H_0$ values, respectively. Given that we have seen little change in the uncalibrated GRBs, but here the change is pronounced as we remove GRBs, it will be interesting to see if this effect is due to the calibration process with respect to Type Ia SN, since a similar emergent dipole has been observed there too \cite{Krishnan:2021jmh}. We leave this question to future work.

\section{Discussion} 
Building on earlier observations in strong lensing time delay \cite{Krishnan:2021dyb} and Type Ia SN \cite{Krishnan:2021jmh}, which already place the significance at $1.7 \sigma$, we find independent evidence in QSOs and GRBs that $H_0$ is larger in the CMB dipole direction or aligned directions. Note that we have simply assumed the flat $\Lambda$CDM model, which \textit{a priori} has no directional preference  and knows nothing about any anisotropy. We focus on $H_0$, since $H_0$ is a universal constant in any FLRW cosmology (for example, see \cite{Krishnan:2020vaf}). For this reason, one expects all trends to be robust across cosmological models, although significance may vary. The use of $H_0$ also allows us to highlight the Hubble constant, the local value of which is the subject of ongoing debate \cite{Riess:2020fzl, Freedman:2021ahq, Pesce:2020xfe, Blakeslee:2021rqi, Kourkchi:2020iyz}, but if our findings hold up, then there is no reason for $H_0$ in an FLRW context to be unique. In essence, ``Hubble tension" may not be the problem it is usually stated to be. 
%even make sense as a problem.

It is certainly worth noting that despite working with proxy constants that are degenerate with $H_0$, FIG. \ref{qso_dipole} negotiates signs to register good agreement across observables with the companion letter \cite{Krishnan:2021jmh}. On the contrary, FIG. \ref{grb_dipole} and FIG. \ref{demianski1} contradict observations in strong lensing, Type Ia SN and QSOs, while FIG. \ref{demianski2} and FIG. \ref{demianski3} show agreement once more. Moreover, just as in the Hubble tension narrative, where one can replace $H_0$ with the absolute magnitude $M_{B}$ of Type Ia SN to define ``$M_{B}$ tension"  \cite{Camarena:2021jlr, Efstathiou:2021ocp}, one has the same freedom with $\beta$ (for QSOs) and $\alpha$ (for GRBs). Given the small sample of GRBs, they simply play an accompanying role, but the significance of the deviation in the QSOs, as defined by our weighted sum (\ref{weighted_sum}), is $\gtrsim 2 \sigma$. Importantly, our lowest redshift QSO and GRB are sufficiently deep in redshift that {within the usual FLRW framework} all the data is expected to be in the CMB frame. Admittedly, since the GRB results are sample dependent, some further work is required to establish whether GRBs are contributing to a coherent narrative. Ultimately, this may have a simple explanation. For example, there is considerable scatter in GRB data sets and when the data is uncalibrated, one has to fit a large number of fitting parameters.

Obviously, QSOs and GRBs are not widely used in cosmology, although this is changing in recent years \cite{Melia:2019nev, Demianski:2019vzl, Khoraminezhad:2020cer, Lindner:2020eez, Zheng:2020fth, Mehrabi:2020zau, Liu:2020bzc, Demianski:2020bva, Hu:2020mzd, Luongo:2020aqw, Cao:2020evz, Speri:2020hwc, Luongo:2020hyk, Muccino:2020gqt, Li:2021onq, Lian:2021tca, Wang:2021hcx}. %, Rezaei:2021qwd}. 
Both have great potential as they probe redshift ranges not accessible by other probes. Nonetheless, they come with caveats, which we have openly discussed in section \ref{sec:qso_caveats}. Moreover, as we have shown here, when working with the latest QSO compilation \cite{Lusso:2020pdb} (see however \cite{Banerjee:2020bjq}), one is implicitly working with a data set that is not only naively at odds with flat $\Lambda$CDM (large $\Omega_{m}$), but also FLRW. As argued, this high redshift window in the late Universe is largely unexplored and even BAO has led to lower values of $D_{A}(z)$ \cite{BOSS:2014hwf, DES:2021esc} that are consistent with the Risaliti-Lusso QSOs. The QSO tendency for larger $\Omega_m$ appears to have a simple explanation (see appendix \ref{sec:QSOHD}). While QSOs see dark energy at lower redshifts, once one moves into the higher redshift sample, the QSOs anchoring the dark energy sector become statistically insignificant and QSOs start to inhabit an Einstein-de Sitter Universe at the level of current precision in the data. This interpretation should also explain similar trends with (binned) high redshift HST SN \cite{Horstmann:2021jjg, Dainotti:2021pqg}, because the connection to SN anchoring the Hubble diagram in the Planck-$\Lambda$CDM Universe has been largely severed. This may explain reported tensions \cite{Risaliti:2018reu}.
%\red{Moreover, recent results in an alternative method to calibrate QSOs hint at the same fall off in $D_{L}(z)$ at higher redshifts \cite{Solomon:2021jml} (see FIG. 6). Finally, HST SN also prefer higher values of $\Omega_{m}$ within flat $\Lambda$CDM \cite{Horstmann:2021jjg, Dainotti:2021pqg}, so deviations from Planck-$\Lambda$CDM may yet emerge in the late Universe just beyond $z \sim 1$. It is clearly imperative to identify a greater number of high redshift SN.} 
 
Given the diverse nature of the observables, e. g. $\log_{10} F_{X}, \log_{10} F_{UV}, E_{\textrm{p,i}}$ and $S_{\textrm{bolo}}$, there is nothing to suggest that we cannot simply combine probabilities following Fisher's method for independent observations to ascertain the probability that statistical fluctuations could explain our observation across strong lensing, Type Ia SN, QSOs and GRBs. Combining strong lensing, Type Ia SN and the more conservative QSO subsample, the significance is at $2.4 \sigma$ (single-sided Gaussian). This increases marginally to $2.7 \sigma$ with the full QSO sample. As we have seen, while analysis based on uncalibrated GRBs remains inconclusive, we are seeing a strong emergent dipole in calibrated GRBs. In particular, we identified 78 calibrated GRBs, where the probability of a fluke was $p = 0.047$. Folding this number into the probabilities, the statistical significance rises to $3 \sigma$. 
%i. e. on par with any discrepancy in $H_0$ \cite{Riess:2020fzl, Freedman:2021ahq, Pesce:2020xfe, Blakeslee:2021rqi, Kourkchi:2020iyz} and something not to be easily dismissed.  
This of course represents an optimistic assessment of the statistical significance pending a further study of GRB data. It should be stressed again that we have not employed any ansatz to guide the data, but simply worked within flat $\Lambda$CDM, so the observation that $H_0$ is higher across i) strong lensing time delay \cite{Krishnan:2021dyb}, ii) Type Ia SN \cite{Krishnan:2021jmh}, iii) QSOs and iv) GRBs, surely must have some interesting physical explanation. Obviously, a Universe with an anisotropic Hubble expansion is the simplest interpretation and this claim can be tested by any competitive cosmological data set going forward \footnote{Interestingly, a recent dark siren $H_0$ determination \cite{Palmese:2021mjm} is consistent with our narrative when the PDFs in Figure 4 are decomposed in hemispheres. More concretely, GW190412 drives larger $H_0$ values and it is localised in the hemisphere with the CMB dipole. It will be interesting to see how this develops.}. We have a prediction and as we gather more and better quality data the claim can either be true or false; our prediction has a good prospect to be verified but community engagement is required.

We close by briefly discussing the theoretical ramifications of our results. If further data confirms our findings, the simplest explanation may be that we have a preferred direction, aligned with the CMB dipole, in the Universe. That is, going to CMB rest-frame we see an anisotropic background. Homogeneous but anisotropic cosmologies are classified by Bianchi models. Since flat $\Lambda$CDM is apparently already a good approximation to the Universe, such Bianchi models should be anisotropic deviations from the flat $\Lambda$CDM. Moreover, our findings require the presence of a preferred ``dipole" direction, which may be found in specific Bianchi models, such as the ``tilted cosmology" of King and Ellis \cite{Tilted-Cosmology-1,Tilted-Cosmology-2}. Going beyond FLRW, one should revisit all the cosmological analyses and inferences, and write a new chapter in the cosmology book.

\section*{Acknowledgements}
We thank Stephen Appleby, Narayan Khadka, Bharat Ratra and Jian-Min Wang for helpful discussions. We are grateful to Mohamed Rameez, Jenny Wagner and an anonymous referee for comments on the draft. We especially thank Chethan Krishnan and Roya Mohayaee for ongoing discussions on related topics. E\'OC is funded by the National Research Foundation of Korea (NRF-2020R1A2C1102899). MMShJ would like to acknowledge SarAmadan
grant No. ISEF/M/400122. LY was supported
by the National Research Foundation of Korea funded through the CQUeST of Sogang University (NRF2020R1A6A1A03047877). OL and MM acknowledge the Ministry of Education and Science of the Republic of Kazakhstan, Grant: IRN AP08052311.

\appendix 

\section{Comments on QSO Hubble Diagram} 
\label{sec:QSOHD}
The redshift limitations of Type Ia SN are clear and for this reason there is understandable interest in emerging cosmological probes in the late Universe (see \cite{Moresco:2022phi} for a recent review). QSOs are part of the mix, but they challenge a number of preconceptions based on familiarity with SN. First, it is not clear if QSOs can ever rival SN as standardisable candles, but since scaling relations are commonplace in astrophysics, it is not so surprising that UV and X-ray fluxes satisfy a non-linear relation in QSOs (see Fig. 3 of Ref. \cite{Risaliti:2015zla}). As remarked in the text, the slope $\gamma$ of this relation is robust to binning \cite{Vignali:2002ct, Just:2007se, Lusso:2009nq, Salvestrini:2019thn, Bisogni:2021hue}. Assuming the usual flux-luminosity relation, this implies $L_{X} = \beta(z) L^{\gamma}_{UV}$, where the X-ray $L_{X}$ and UV luminosities $L_{UV}$ are assumed to be intrinsic to any given QSO. The assumption of Risaliti \& Lusso is that $\beta$ does not evolve with redshift, so that the relation is completely intrinsic to QSO physics. That being said, as touched up in the text in section \ref{sec:qso_caveats}, there may be some evolution of $(\beta, \gamma)$ with redshift. Note, however that ($\beta, \gamma)$ are correlated, so \textit{a priori} it is not clear if $\beta$ or $\gamma$ evolves. It should be noted again that the Risaliti-Lusso QSOs are still in their infancy and remain a powerful proposal that spans a redshift range considerably greater than SN. Moreover, since their inception, even SN have been corrected for i) colour ii) shape and iii) mass of the host galaxy, so it is worth bearing in mind again that standardisable candles are living, breathing working assumptions that may require corrections. 

\begin{table}[htb]
\centering 
\begin{tabular}{c|c|c|c}
 \rule{0pt}{3ex} $z_{\textrm{max}}$ & $ \Omega_{m}$ & $ \beta$ & $\gamma$ \\
\hline 
\rule{0pt}{3ex} \multirow{2}{*}{$0.7$ (398 QSOs)} & $0.266$ & $6.601$  & $0.670$ \\
\rule{0pt}{3ex}  & $0.428^{+0.329}_{-0.268}$ & $6.586^{+0.842}_{-0.796}$ & $0.670^{+0.027}_{-0.026}$  \\
\hline
\rule{0pt}{3ex} \multirow{2}{*}{$0.8$ (543 QSOs)} & $0.418$ & $7.162$  & $0.652$ \\
\rule{0pt}{3ex}  & $0.511^{+0.305}_{-0.275}$ & $7.162^{+0.715}_{-0.712}$ & $0.651^{+0.023}_{-0.023}$  \\
\hline
\rule{0pt}{3ex} \multirow{2}{*}{$0.9$ (678 QSOs)} & $0.592$ & $7.736$  & $0.633$ \\
\rule{0pt}{3ex}  & $0.601^{+0.248}_{-0.250}$ & $7.709^{+0.662}_{-0.679}$ & $0.633^{+0.022}_{-0.021}$  \\
\hline
\rule{0pt}{3ex} \multirow{2}{*}{$1$ (826 QSOs)} & $0.953$ & $7.921$  & $0.626$ \\
\rule{0pt}{3ex}  & $0.717^{+0.184}_{-0.231}$ & $7.792^{+0.571}_{-0.571}$ & $0.631^{+0.019}_{-0.019}$  \\
\end{tabular}
\caption{Best fit and marginalised inferences of ($\Omega_{m}, \beta, \gamma)$ for QSOs below a maximum redshift $z_{\textrm{max}}$.}
\label{QSOvsZ}
\end{table}

Our goal in this appendix is twofold. First, we address why QSOs prefer larger values of $\Omega_{m} \approx 1$ in contrast to SN, which are ostensibly consistent with Planck-$\Lambda$CDM. Secondly, we will explain why our result should be robust to any hints of evolution in $(\beta, \gamma)$ with redshift (see Fig. \ref{lowz_highz}). Before proceeding, let us remark that as is clear from Fig. 5 of Ref. \cite{Risaliti:2015zla}, Fig. 2 of Ref. \cite{Risaliti:2018reu} and Fig. 9 of Ref. \cite{Lusso:2020pdb}, a redshift range exists in which the QSO Hubble diagram and SN Hubble diagram show good agreement.
%\red{Now, of course, the relation (\ref{X_UV}) is still non-trivial and the assumption of constant $\beta$ may make one uneasy. However, this can checked by cross-calibrating QSOs with SN in a redshift range where we have confidence in SN. The result of this exercise can be seen in Fig. 5 of Ref. \cite{Risaliti:2015zla}, Fig. 2 of Ref. \cite{Risaliti:2018reu} and Fig. 9 of Ref. \cite{Lusso:2020pdb}, and what one notices is that values of $(\beta, \gamma)$ exist so that SN and QSO Hubble diagrams agree, but of course deviate at higher redshifts beyond $z \sim 1.5$. What this says is that there is an inconsistency between the SN and QSO Hubble diagrams, but it only becomes evident at larger redshifts. Note, SN typically become sparse beyond $z=1$, but recently Ref. \cite{Li:2021onq} has employed data reconstruction techniques to calibrate QSOs up to the end of the Pantheon SN sample \cite{Scolnic:2017caz}, $z \sim 2.3$. This extends the earlier efforts of Ref. \cite{Risaliti:2015zla, Risaliti:2018reu} to higher redshifts, and the visibly poorer quality QSO data (see \cite{Risaliti:2015zla, Risaliti:2018reu}) simply adopt distance moduli preferred by SN deeper in redshift. This understandably leads to a reduction in tension between QSOs and SN. Regardless, some discrepancy persists at high redshifts and this deserves more study.}
However, in this paper we have made a conscious decision to treat QSOs and SN independently. This ensures that any directions emerging from QSOs are not biased by SN. As highlighted in the discussion, the role of SN calibration in our GRB results needs to be investigated further. %Nevertheless, we impose a flat prior on matter density within the flat $\Lambda$CDM model in line with physical requirements, $0 < \Omega_{m} < 1$, and while our best-fits saturate the bound, $\Omega_{m} =1$ our goal here is to explain why this may be the case. Teasing out this issue further will be left to a subsequent publication \cite{toappear}, where we will also highlight potential synergies with SN. 
In Table \ref{QSOvsZ} we restrict the redshift range of the QSO sample $0 < z < z_{\textrm{max}}$ and record extremised and marginalised results for the likelihood (\ref{L1}) with increasing redshift. 

\begin{figure}[htb]
\centering
  \includegraphics[width=80mm]{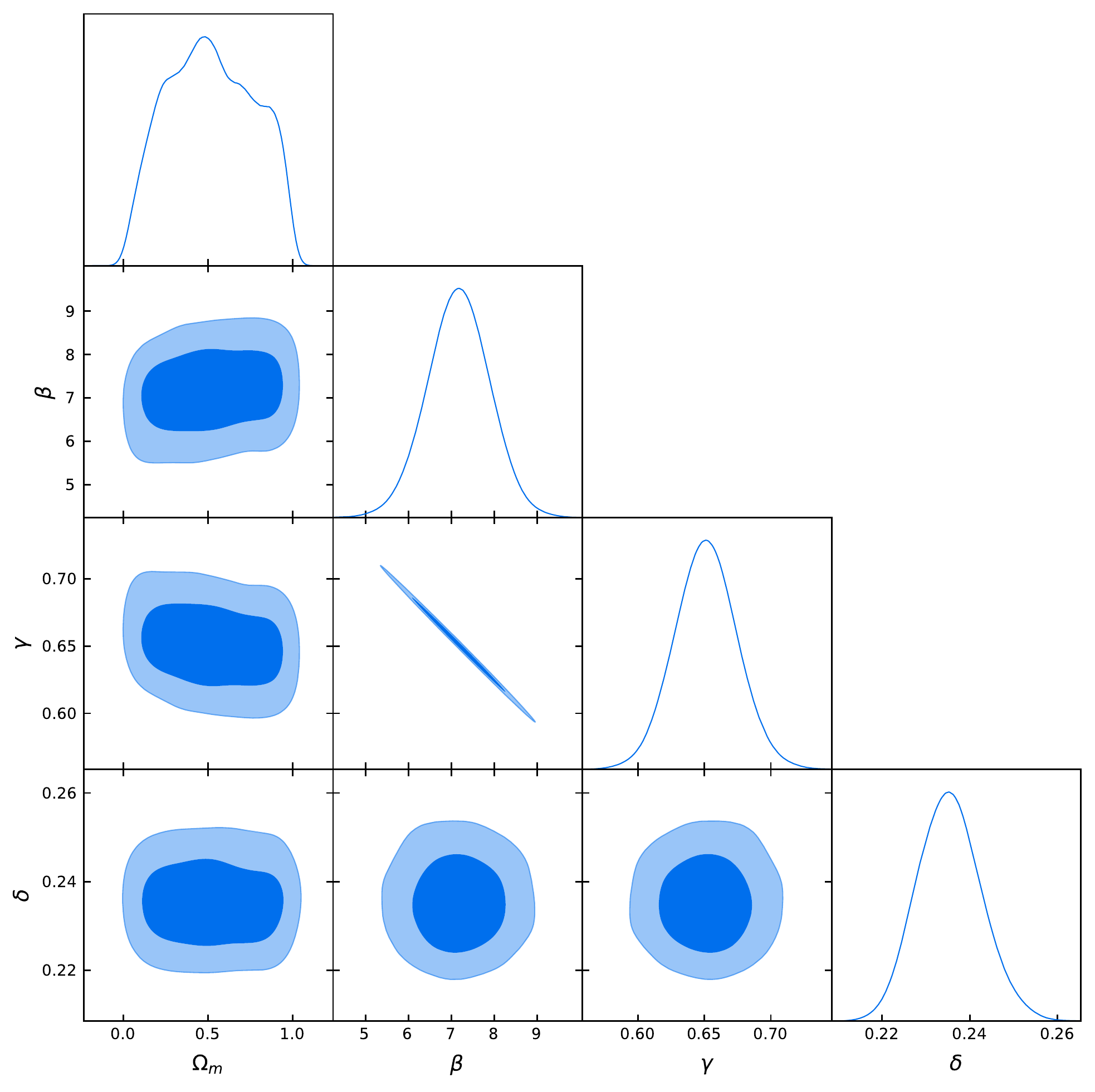}
\caption{MCMC exploration of the likelihood (\ref{L1}) in the redshift range $0 < z < 0.8$.}
\label{lowzQSO}
\end{figure}

Remarkably, we see that QSOs recover a Planck-$\Lambda$CDM, in agreement with SN when fitted over similar redshift ranges. This can be done without cross-calibration. However, in contrast to SN, which are weighted to low redshifts and therefore tightly anchored in the dark energy regime, Risaliti-Lusso QSOs become more numerous at higher redshifts. This means as we increase $z_{\textrm{max}}$, higher redshift QSOs, which struggle to distinguish Einstein-de Sitter \footnote{a spatially flat FLRW Universe with matter} and Planck-$\Lambda$CDM Universes come to dominate the statistically less significant QSOs anchoring the Hubble diagram in the dark energy dominated regime. One finds the same feature in SN by steadily removing the anchoring lower redshift SN \cite{Horstmann:2021jjg, Dainotti:2021pqg}. Incidentally, the same trend is evident in the new Pantheon+ sample (see FIG. 16 of Ref. \cite{Brout:2022vxf}). In FIG. \ref{lowzQSO}, we document the output from an MCMC exploration of the QSO likelihood (\ref{L1}) with $z_{\textrm{max}} = 0.8$. This not only confirms some of the content of Table \ref{QSOvsZ}, but it shows us that $(\beta, \gamma)$ are largely insensitive to the value of $\Omega_{m}$, and as a result, the changes in $\Omega_{m}$ have little impact on $\beta$. Thus, even if one fixes $\Omega_{m}$ to its Planck value through a prior or incorporating SN, one expects similar results.

\begin{figure}[htb]
\centering
  \includegraphics[width=80mm]{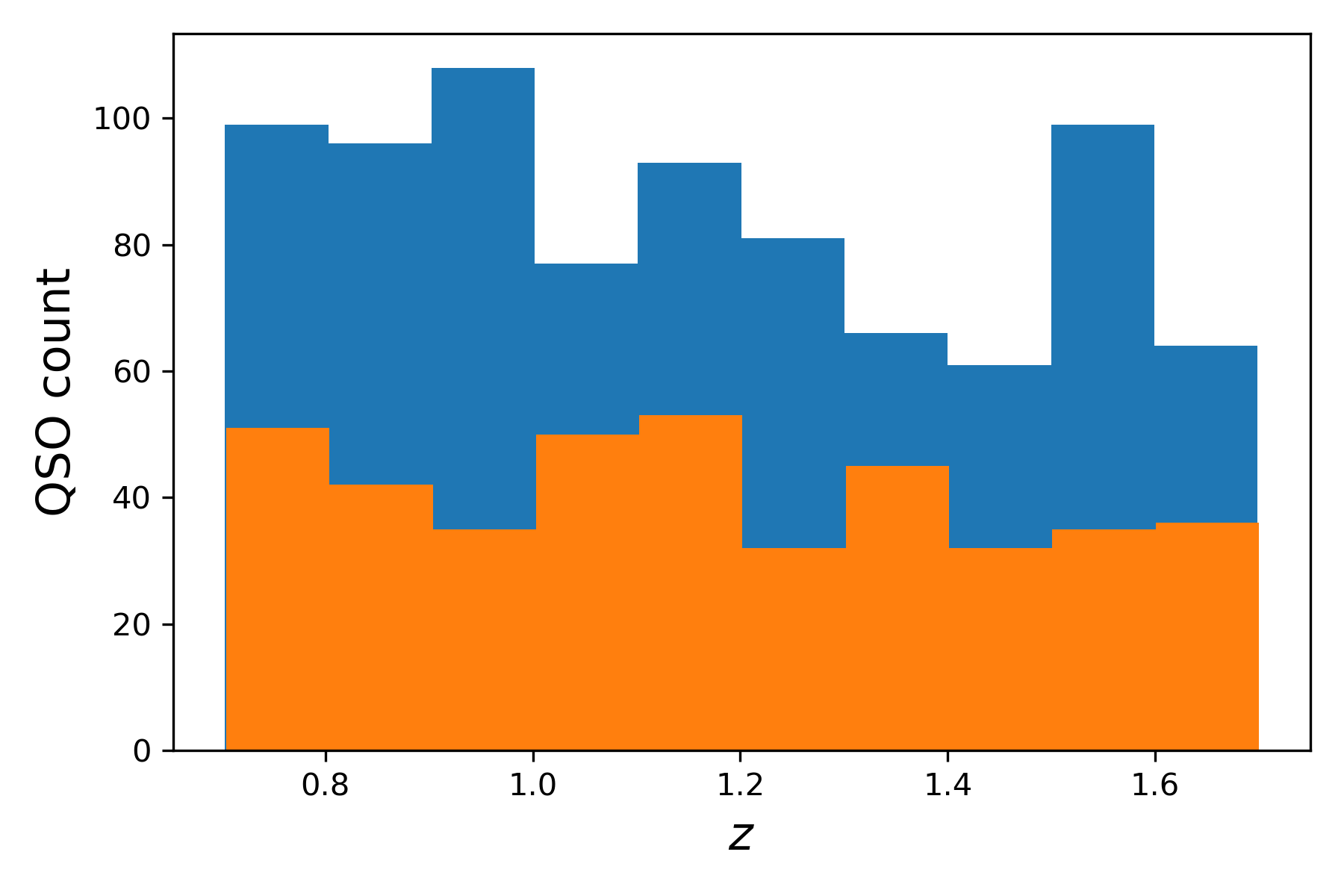}
\caption{QSO counts in the CMB dipole direction (blue) versus opposite direction (orange) in the redshift range $0.7 < z < 1.7$.}
\label{QSOcounts_redshift}
\end{figure}

Now, returning to Table \ref{QSOvsZ}, one sees that there is some evolution of ($\beta, \gamma$) with $z_{\textrm{max}}$, but evidently this evolution becomes less pronounced as one moves to higher redshifts. It is timely to refer the reader back to Ref. \cite{Lusso:2020pdb}, where it is explained that QSOs below $z=0.7$ may be less reliable owing to extrapolations of UV fluxes from the optical. Once we remove these QSOs, we can expect smaller displacements in $\beta$, although as is clear from FIG. \ref{lowz_highz}, this can lead to sizable shifts over extended redshift ranges. Nevertheless, provided the QSOs are distributed fairly uniformly in the CMB dipole direction and the opposite direction, it is hard to imagine that this (small) evolution in redshift could mimic an orientational dependence of $(\beta, \gamma)$. In FIG. \ref{QSOcounts_redshift} we confirm that the QSOs are uniformly distributed in the CMB dipole hemisphere (blue) and opposite hemisphere (orange). More precisely, QSO counts in the CMB dipole hemisphere show a marginal fall off in count with redshift, but this simply counteracts any increase in $\beta$ in the redshift range $0.7 < z < 1.7$. In short, provided the QSOs are fairly uniformly distributed with redshift, one does not expect any bias from evolution in $\beta$. 

Let us summarise our findings. As we have shown, both SN and QSOs recover the Planck-$\Lambda$CDM Universe in redshift ranges where SN are numerous. Moreover, this does not require the QSOs to be calibrated. That being said, QSOs are biased to higher redshifts where any signature of dark energy is weaker. For this reason, one expects matter density to be closer to unity. One can see the same feature in SN by removing the lower redshift SN that anchor the Hubble diagram in the dark energy regime \cite{Colgain:2022nlb}. Separately, we have seen that there may be some evolution in $(\beta, \gamma)$ with redshift, but since our QSO counts are pretty uniform with redshift, this is not expected to impact results. Obviously, correcting for this evolution is important for developing the Risaliti-Lusso QSOs further \cite{Dainotti:2022rfz}, but it is worth stressing once again that the anisotropy we find is not only consistent with a similar observation in SN \cite{Krishnan:2021jmh}, but also claims of an excess in the QSO cosmic dipole \cite{Secrest:2020has}.

\section{Data}
We record the GRB data compiled earlier in \cite{Khadka:2021dsz} in the below table. The QSO data \cite{Lusso:2020pdb} can be downloaded from https://vizier.u-strasbg.fr/viz-bin/VizieR?-source=J/A+A/642/A150. 

%\begin{center}
\begin{longtable}{|l|l|l|l|l|l|}
\caption{The GRB data points} \label{tab:long} \\
			\hline \multicolumn{1}{|c|}{\textbf{Name}} & \multicolumn{1}{c|}{\textbf{Redshift $z$}} & \multicolumn{1}{c|}{\textbf{$E_p$}} & \multicolumn{1}{c|}{\textbf{$S_{bolo}$}}& \multicolumn{1}{c|}{\textbf{RA}}& \multicolumn{1}{c|}{\textbf{DEC}}\\ \hline 
			\endfirsthead
			
			\multicolumn{6}{c}%
			{{}} \\
			\hline \multicolumn{1}{|c|}{\textbf{Name}} & \multicolumn{1}{c|}{\textbf{redshift $z$}} & \multicolumn{1}{c|}{\textbf{$E_p$}} & \multicolumn{1}{c|}{\textbf{$S_{bolo}$}}& \multicolumn{1}{c|}{\textbf{RA}}& \multicolumn{1}{c|}{\textbf{DEC}}\\ \hline 
			\endhead
			
%			\hline \multicolumn{6}{|r|}{{Continued on next page}} \\   \hline
			\hline \multicolumn{6}{r}{} 
			\endfoot
			
			\hline
			\endlastfoot
 
		080916C & 4.35 &   6953.87 $\pm$ 1188.77 &  10.40 $\pm$ 0.24   &  119.88& -57\\
		090323  & 3.57 &   2060.09 $\pm$ 138.07  &  15.76 $\pm$   0.39   &  190.69& 17\\
		090328  & 0.736 &  1221.71 $\pm$ 81.87  &   7.99  $\pm$ 0.20    &  90.87& -42\\
		090424  & 0.544 &  236.91 $\pm$ 4.55    &   5.72  $\pm$ 0.09    &  189.54& 17\\
		090902B & 1.822 &  2146.57 $\pm$ 21.71  &   39.05 $\pm$   0.22   &  265.00 &27\\
		090926A & 2.1062 & 868.63 $\pm$ 13.85   &   17.90 $\pm$   0.13   &  353.25 &-39\\
		091003A & 0.8969 & 857.81 $\pm$ 33.08   &   4.43  $\pm$  0.08    &  251.50 &37\\
		091127  & 0.49   & 60.32 $\pm$ 1.93      &  2.25  $\pm$  0.04    &  36.57& -19\\
		091208B & 1.063  & 202.63 $\pm$ 20.10    &  0.75  $\pm$  0.04    &  29.41 &17\\
		100414A & 1.368  & 1370.82 $\pm$ 27.68    & 11.88 $\pm$   0.16   &  183.62& 20\\
		100728A & 1.567  & 797.62  $\pm$ 18.05   &  11.74 $\pm$  0.17   &       77.07& -14
\\		110721A & 3.512  & 8675.78 $\pm$ 852.66 &  6.14   $\pm$  0.09    &     333.40 &-39
\\		120624B & 2.2    & 1214.47 $\pm$ 26.24 &   20.49  $\pm$  0.25   &      170.94& 9
\\		130427A & 0.3399 & 294.25  $\pm$ 5.86 &     31.72 $\pm$  0.20   &        173.14 &28
\\		130518A & 2.49   & 1601.40 $\pm$ 32.19&    11.40  $\pm$  0.11   &      355.81& 48
\\		131108A & 2.40   & 1163.20 $\pm$ 28.54&    4.85   $\pm$ 0.05    &      156.47& 10
\\		131231A & 0.6439 & 370.15  $\pm$ 4.97&      17.42 $\pm$   0.12   &        10.58& -2
\\		141028A & 2.33   & 1320.18 $\pm$ 50.90&    4.89   $\pm$ 0.06    &      322.70 &0
\\		150314A & 1.758  & 985.66  $\pm$ 13.20 &    9.20  $\pm$  0.12    &       126.66& 64
\\		150403A & 2.06   & 2428.51 $\pm$ 160.80&   8.10   $\pm$  0.17    &     311.50& -63
\\		150514A & 0.807  & 137.84  $\pm$ 14.93 &    0.71  $\pm$  0.03    &       74.85& -61
\\		160509A & 1.17   & 19334.10 $\pm$  652.25&  49.91 $\pm$   1.36   &    310.10 &76
\\		160625B & 1.406  & 1546.86  $\pm$ 37.25&    83.54 $\pm$   1.16   &      308.27& 7
\\		170214A & 2.53   & 2119.788  $\pm$ 119.06&  22.40 $\pm$   0.29   &    256.33& -2
\\		170405A & 3.51   & 1424.42  $\pm$  35.24 &   9.24 $\pm$   0.09    &      219.81& -25
\\		971214  & 3.42   & 685.0  $\pm$  133.0 &     0.87 $\pm$   0.11    &        179.13 &65
\\		990123  & 1.6    & 1724.0  $\pm$  466.0 &    35.80 $\pm$   5.80   &       231.37 &45
\\		990510  & 1.619  & 423.0 $\pm$   42.0     &  2.60 $\pm$   0.40    &         204.53 &-81
\\		000131  & 4.5    & 987.0 $\pm$   416.0&      4.70 $\pm$   0.80    &        93.39&-52
\\		000926  & 2.07   & 310.0  $\pm$  20.0 &      2.60 $\pm$   0.60    &         256.06 &52  
\\		010222  & 1.48   & 766.00 $\pm$   30.0 &     14.6 $\pm$   1.50    &        223.05& 43
\\		011211  & 2.14   & 186.0  $\pm$  24.0&       0.50 $\pm$   0.06    &         168.82& -22
\\		020124  & 3.2    & 448.0  $\pm$  148.0 &     1.20 $\pm$   0.10    &        143.28&-12
\\		021004  & 2.3    &266.0  $\pm$  117.0 &      0.27 $\pm$   0.04    &        6.73 &19
\\		030226  & 1.98   & 289.0 $\pm$   66.0 &      1.30 $\pm$   0.10    &         173.27& 26
\\		030323  & 3.37   & 270.0  $\pm$  113.0 &     0.12 $\pm$   0.04    &        166.53 &-22
\\		030328  & 1.52   & 328.00  $\pm$  55.0&      6.40 $\pm$   0.60    &        182.69 &-9
\\		030429  & 2.65   & 128.0  $\pm$  26.0 &      0.14 $\pm$   0.02    &         183.28 &-21
\\		040912  & 1.563  & 44.00 $\pm$   33.0&    0.21    $\pm$ 0.06       &      359.00 &-1
\\		050318  & 1.44   & 115.00 $\pm$   25.0&   0.42    $\pm$  0.03       &     49.68& -46
\\		050401  & 2.9    & 467.0  $\pm$  110.0& 1.90      $\pm$  0.40         &   247.88& 2
\\		050603  & 2.821  & 1333.0 $\pm$   107.0& 3.50     $\pm$  0.20        &   39.98& -25
\\		050820  & 2.612  & 1325.0 $\pm$   277.0& 6.40  $\pm$  0.50        &   337.40 &20
\\		050904  & 6.29   & 3178   $\pm$ 1094.0& 2.00   $\pm$ 0.20         &   13.67&14
\\		050922C & 2.198  & 415.0  $\pm$  111.0& 0.47   $\pm$ 0.16         &   317.39 &-9
\\		051109A & 2.346  & 539.0  $\pm$  200.0& 0.51   $\pm$ 0.05         &   330.25 &41
\\		060115  & 3.53   & 285.0  $\pm$  34.0 &0.25    $\pm$ 0.04          &   54.05& 17
\\		060124  & 2.296  & 784.0  $\pm$  285.0& 3.40   $\pm$ 0.50         &   77.04& 70
\\		060206  & 4.048  & 394.0  $\pm$  46.0 &0.14    $\pm$ 0.03          &   202.95 &35
\\		060418  & 1.489  & 572.00 $\pm$   143.0& 2.30  $\pm$  0.50        &   236.42 &-4
\\		060526  & 3.21   & 105.0  $\pm$  21.0& 0.12    $\pm$ 0.06          &   232.85 &0
\\		060707  & 3.425  & 279.0  $\pm$  28.0 &0.23    $\pm$  0.04          &   357.08& -18
\\		060908  & 2.43   & 514.0  $\pm$  102.0 &0.73   $\pm$  0.07         &   31.83& 0
\\		060927  & 5.6    & 475.0  $\pm$  47.0 &0.27    $\pm$ 0.04          &   329.55 &5
\\		070125  & 1.547 &  934.00  $\pm$  148.0& 13.30 $\pm$   1.30       &   117.85 &31
\\		071003  & 1.604 &  2077  $\pm$  286 &5.32      $\pm$ 0.590           &   301.85& 11
\\		071020  & 2.145 &  1013.0 $\pm$   160.0 &0.87  $\pm$   0.40        &   119.66 &33
\\		080319C & 1.95  &  906.0  $\pm$  272.0 &1.50   $\pm$  0.30         &   258.98 &55
\\		080413  & 2.433 &  584.0 $\pm$   180.0 &0.56   $\pm$  0.14         &   287.29& -28
\\		080514B & 1.8   &  627.0 $\pm$   65.0 &  2.027 $\pm$   0.48       &     322.82& -1
\\		080603B & 2.69  &  376.0 $\pm$   100.0 &0.64   $\pm$ 0.058        &   176.53 &68
\\		080605  & 1.6398&  650.0 $\pm$   55.0 &3.40    $\pm$ 0.28          &   262.13& 4
\\		080607  & 3.036 &  1691.0 $\pm$   226.0& 8.96  $\pm$  0.48        &   194.97& 16
\\		080721  & 2.591 &  1741.0 $\pm$   227.0 &7.86  $\pm$  1.37        &   224.47 &-12
\\		080810  & 3.35  &  1470.0 $\pm$   180.0 &1.82  $\pm$  0.20        &   356.78 &0
\\		080913  & 6.695 &  710.0 $\pm$   350.0 &0.12   $\pm$ 0.035        &   65.73 &-25
\\		081008  & 1.9685&  261.0 $\pm$   52.0 &0.96    $\pm$  0.09          &   279.99 &-57
\\		081028  & 3.038 &  234.0 $\pm$   93.0 &0.81    $\pm$  0.095         &   121.89& 2
\\		081118  & 2.58  &  147.0 $\pm$   14.0 &0.27    $\pm$  0.057         &   82.59& -43
\\		081121  & 2.512 &  47.23 $\pm$   1.08 &1.71    $\pm$  0.33          &   89.26&-61
\\		081222  & 2.77  &  505.0 $\pm$   34.0 &1.67    $\pm$ 0.17          &   22.75 &-34
\\		090102  & 1.547 &  1149.00 $\pm$   166.0& 3.48 $\pm$   0.63       &   128.26 &33
\\		090418  & 1.608 &  1567  $\pm$  384 &2.35     $\pm$  0.59            &   269.33& 33
\\		090423  & 8.2   &  491.0  $\pm$  200.0& 0.12  $\pm$   0.032        &   148.90 &18
\\		090516  & 4.109 &  971.0  $\pm$  390.0 &1.96  $\pm$  0.38         &   138.27& -12
\\		090715B & 3.0   &  536.0  $\pm$  172.0 &1.09  $\pm$  0.17         &   251.35& 45
\\		090812  & 2.452 &  2000.0  $\pm$  700.0& 3.08 $\pm$   0.53        &   353.20 &-11
\\		091020  & 1.71  &  280.0   $\pm$ 190.0 &0.11  $\pm$  0.034        &   175.72& 51
\\		091029  & 2.752 &  230.0   $\pm$ 66.0  &0.47  $\pm$  0.044        &    60.18& -56
\\		100413  & 3.9   &  1783.60  $\pm$  374.85& 2.36 $\pm$   0.77      &  266.00 &16 
\\		100621  & 0.54  &  146.49  $\pm$  23.9 &5.75  $\pm$  0.64         &  315.25& -51
\\		100704  & 3.6   &  809.60  $\pm$  135.70& 0.70 $\pm$   0.07       &  133.50 &-24
\\		100814  & 1.44  &  312.32  $\pm$  48.8 &1.39  $\pm$  0.23         &  22.25 &-18
\\		100906  & 1.73  &  387.23  $\pm$  244.07& 3.56 $\pm$   0.55       &  28.50 &56
\\		110205  & 2.22  &  740.60  $\pm$  322.0 &3.32  $\pm$  0.68        &  164.50& 68
\\		110213  & 1.46  &  223.86  $\pm$  70.11 &1.55  $\pm$  0.23        &  43.00 &49
\\		110422  & 1.77  &  421.04  $\pm$  13.85 &9.32  $\pm$  0.02        &  111.75& 75
\\		110503  & 1.61  &  572.25  $\pm$  50.95 &2.76  $\pm$  0.21        &  132.75& 52
\\		110715  & 0.82  &  218.40  $\pm$  20.93 &2.73  $\pm$  0.24        &  237.50 &-46
\\		110731  & 2.83  &  1164.32 $\pm$   49.79 &2.51 $\pm$  0.01        & 280.50& -29
\\		110818  & 3.36  &  1117.47 $\pm$   241.11 &1.05 $\pm$   0.08      &  317.25 &-64
\\		111008  & 5.0   &  894.00  $\pm$  240.0 &1.06 $\pm$   0.11        &  60.25 &-33
\\		111107  & 2.89  &  420.44  $\pm$  124.58 &0.18  $\pm$  0.03       &  129.25 &-67
\\		111209  & 0.68  &  519.87  $\pm$  88.88 &69.47  $\pm$  8.72       &  14.25& -47
\\		120119  & 1.73  &  417.38  $\pm$  54.56 &4.62   $\pm$ 0.59        &  120.00 &-9
\\		120326  & 1.8   &  129.97  $\pm$  10.27& 0.44  $\pm$  0.02        &  273.75& 69
\\		120724  & 1.48  &  68.45   $\pm$ 18.60 &0.15   $\pm$ 0.02         &  245.00 &4
\\		120802  & 3.8   &  274.33  $\pm$  93.04 &0.43  $\pm$  0.07        &  44.75& 14
\\		120811C & 2.67  &  157.49  $\pm$  20.92 &0.74  $\pm$  0.07        &  199.71 &62
\\		120909  & 3.93  &  1651.55 $\pm$   123.25& 2.69 $\pm$   0.23      &  275.50 &-59
\\		120922  & 3.1   &  156.62  $\pm$  0.04 &1.59   $\pm$ 0.18         &  234.75& -20
\\		121128  & 2.2   &  243.20  $\pm$  12.8 &0.87   $\pm$ 0.07         &  300.50& 54
\\		130215  & 0.6   &  247.54  $\pm$  100.61& 4.84  $\pm$  0.12       &  43.25 &13
\\		130408  & 3.76  &  1003.94 $\pm$   137.98& 0.99  $\pm$  0.17      &  134.25& -32
\\		130420A & 1.3   &  128.63  $\pm$  6.89 &1.73 $\pm$  0.06          &  196.10 &59
\\		130505  & 2.27  &  2063.37 $\pm$   101.37& 4.56  $\pm$  0.09      &  137.00& 17
\\		130514  & 3.6   &  496.80  $\pm$ 151.8 &1.88  $\pm$  0.25         &  296.25 &-8
\\		130606  & 5.91  &  2031.54 $\pm$   483.7& 0.49  $\pm$  0.09       &  249.25 &30 
\\		130610  & 2.09  &  911.83  $\pm$  132.65 &0.82  $\pm$  0.05       &  224.25 &28
\\		130612  & 2.01  &  186.07  $\pm$  31.56 &0.08   $\pm$ 0.01        &  259.75 &-17
\\		130701A & 1.16  &  191.80  $\pm$  8.62& 0.46  $\pm$  0.04         &  224.42 &28
\\		130831A & 0.48  &  81.35   $\pm$ 5.92& 1.29  $\pm$  0.07          &  358.65 &29
\\		130907A & 1.24  &  881.77  $\pm$  24.62& 75.21 $\pm$   4.76       &  215.88& 46
\\		131030A & 1.29  &  405.86  $\pm$  22.93& 1.05  $\pm$  0.10        &  345.08 &-5
\\		131105A & 1.69  &  547.68  $\pm$  83.53& 4.75  $\pm$  0.16        &  71.01 &-63
\\		131117A & 4.04  &  221.85  $\pm$  37.31& 0.05  $\pm$  0.01        &  332.34& -32
\\		140206A & 2.73  &  447.60  $\pm$  22.38& 1.69  $\pm$  0.03        &  145.35 &67
\\		140213A & 1.21  &  176.61  $\pm$  4.42& 2.53   $\pm$ 0.04         &  105.22& -73\\
\end{longtable}
%\end{center}

\end{document}